\documentclass{article}%
\usepackage{amsfonts}
\usepackage{amsmath}
\usepackage{amssymb}
\usepackage{graphicx}%
\providecommand{\U}[1]{\protect\rule{.1in}{.1in}}

\begin{document}

\title{Maxwell Superalgebras and Abelian Semigroup Expansion }
\author{P. K. Concha$^{1,2,3}$, E. K. Rodr\'{\i}guez$^{1,2,3}$\\\ $^{1}${\small {\textit{Departamento de F\'{\i}sica, Universidad de
Concepci\'{o}n,}} }\\{\small {\textit{Casilla 160-C, Concepci\'{o}n, Chile.}} }\\\ $^{2}${\small {\textit{Dipartimento di Scienza Applicata e Tecnologia
(DISAT),}} }\\{\small {\textit{Politecnico di Torino, Corso Duca degli Abruzzi 24,}}}\\{\small {\textit{I-10129 Torino, Italia.}} }\\\ $^{3}${\small {\textit{Istituto Nazionale di Fisica Nucleare (INFN) Sezione
di Torino,}} }\\{\small {\textit{Via Pietro Giuria, 1 10125 Torino, Italia.}}}}
\maketitle

\begin{abstract}
The Abelian semigroup expansion is a powerful and simple method to derive new
Lie algebras from a given one. \ Recently it was shown that the $S$-expansion
of $\mathfrak{so}\left(  3,2\right)  $ leads us to the Maxwell algebra
$\mathcal{M}$. \ In this paper we extend this result to superalgebras, by
proving that different choices of abelian semigroups $S$ lead to interesting
$D=4$ Maxwell Superalgebras. \ \ \ In particular, the minimal Maxwell
superalgebra $s\mathcal{M}$ and the $N$-extended Maxwell superalgebra
$s\mathcal{M}^{\left(  N\right)  }$ recently found by the Maurer Cartan
expansion procedure, are derived alternatively as an $S$-expansion of
$\mathfrak{osp}\left(  4|N\right)  $. \ Moreover we show that new minimal
Maxwell superalgebras type $s\mathcal{M}_{m+2}$ and their $N$-extended
generalization can be obtained using the $S$-expansion procedure.

\end{abstract}

\section{Introduction}

\bigskip

\qquad The derivation of new Lie algebras from a given one is particularly
interesting in Physics since it allows us to find new physical theories from
an already known. \ In fact, an important example consists in obtaining the
Poincar\'{e} algebra from the Galileo algebra using a deformation process
which can be seen as an algebraic prediction of Relativity. \ At the present,
there are at least four different ways to relate new Lie algebras. In
particular, the expansion method lead to higher dimensional new Lie algebra
from a given one. \ The expansion procedure was first introduced by Hadsuda
and Sakaguchi in Ref. \cite{1} in the context of $AdS$ superstring. \ An
interesting expansion method was proposed by Azcarraga, Izquierdo, Pic\'{o}n
and Varela in Ref. \cite{2} and subsequently developed in Refs. \cite{3, 4}.
\ This expansion method known as Maurer-Cartan (MC) forms power-series
expansion consists in rescaling some group parameters by a factor $\lambda$,
and then apply an expansion as a power series in $\lambda$. \ This series is
truncated in a way that the Maurer-Cartan equations of the new algebra are satisfied.

Another expansion method was proposed by Izaurieta, Rodriguez and Salgado in
Ref. \cite{5} which is based on operations performed directly on the algebra
generators. \ This method consists in combining the inner multiplication law
of a semigroup $S$ with the structure constants of a Lie algebra
$\mathfrak{g}$ in order to define the Lie bracket of a new algebra
$\mathfrak{G}=S\times\mathfrak{g}$. \ This Abelian Semigroup expansion
procedure can reproduce all Maurer-Cartan forms power series expansion for a
particular choice of a semigroup $S$. \ Interestingly, different choices of
the semigroup yield to new expanded Lie algebras that cannot be obtained by
the MC expansion.

An important property of the $S$-expansion procedure is that it provides us
with an invariant tensor for the $S$-expanded algebra in terms of an invariant
tensor for the original algebra. \ This is particularly useful in order to
construct Chern-Simons and Born-Infeld like actions.

Some examples of algebras obtained as an $S$-expansion can be found in Refs.
\cite{5, 6} where the D'auria-Fr\'{e} superalgebra introduced originally in
Ref. \cite{7} and the $M$ algebra are derived alternatively as an
$S$-expansion of $\mathfrak{osp}\left(  32|1\right)  $. \ Subsequently, in
Refs. \cite{8, 9} it was shown that standard odd-dimensional General
Relativity can be obtained from Chern--Simons gravity theory for a certain Lie
algebra $\mathfrak{B}_{m}$ and recently it was found that standard
even-dimensional General Relativity emerges as a limit of a Born-Infeld like
theory invariant under a certain subalgebra of the Lie algebra $\mathfrak{B}%
_{m}$ \cite{9, 11}. \ Very recently it was found that the so-called
$\mathfrak{B}_{m}$ Lie algebra correspond to the Maxwell algebras type%
\footnote{alternatively known as generalized Poincar\'{e} algebra.}
$\mathcal{M}_{m}$ \cite{10}. \ These Maxwell algebras type $\mathcal{M}_{m}$
can be obtained as an $S$-expansion fo the $AdS$ algebra using $S_{E}^{\left(
N\right)  }=\left\{  \lambda_{\alpha}\right\}  _{\alpha=0}^{N+1}$ as an
abelian semigroup.

The Maxwell algebra has been extensively studied in Refs. \cite{12, 13, 14,
15, 16, 17, 18, 19, 20, 21, 22}.\ \ This algebra describes the symmetries of a
particle moving in a background in the presence of a constant electromagnetic
field \cite{12}. \ Interestingly, in Ref. \cite{17} it was shown that a
Maxwell extension of Einstein gravity leads to a generalized cosmological
term. \ Furthermore, it was introduced in Ref. \cite{15} the minimal $D=4$
Maxwell superalgebra $s\mathcal{M}$ which contains the Maxwell algebra as its
bosonic subalgebra. \ In Ref. \cite{19} the Maurer-Cartan expansion was used
in order to obtain the minimal Maxwell superalgebra and its $N$-extended
generalization from the $\mathfrak{osp}\left(  4|N\right)  $
superalgebra.\ This Maxwell superalgebra may be used to obtain the minimal
$D=4$ pure supergravity from the $2$-form curvature associated to
$s\mathcal{M}$ \cite{21}.

The purpose of this work is\ to show that the abelian semigroup expansion is
an alternative expansion method to obtain the Maxwell superalgebra and the
$N-$extended cases. \ In this way we show that the results of Ref. \cite{19}
can be derived alternatively as an $S$-expansion of the $\mathfrak{osp}\left(
4|N\right)  $ superalgebra choosing appropriate semigroups. \ In particular,
the minimal Maxwell superalgebra $s\mathcal{M}$ is obtained as an
$S$-expansion setting a generator equals to zero. \ We finally generalize
these results proposing new Maxwell superalgebras namely, the minimal Maxwell
superalgebras type $s\mathcal{M}_{m+2}$ and the $N$-extended superalgebras
$s\mathcal{M}_{m+2}^{\left(  N\right)  }$ which can be derived from the
$\mathfrak{osp}\left(  4|N\right)  $ superalgebra using the $S$-expansion procedure.

The new $s\mathcal{M}_{m+2}$ superalgebras introduced here are obtained by
applying the $S$-expansion to the $\mathfrak{osp}\left(  4|1\right)  $
superalgebra and can be seen as supersymmetric extensions of the Maxwell
algebras type introduced in Ref. \cite{9}. \ Unlike the Maxwell superalgebra
$s\mathcal{M}$ these new superalgebras involve a larger number of extra
fermionic generators depending on the value of $m$. \ \ Furthermore, these
superalgebras could be used to construct dynamical actions in $D=4$ leading to
standard supergravity, in a very similar way to the bosonic case considered in
Ref. \cite{9}.

This work is organized as follows: In section II, we briefly review some
aspects of the $S$-expansion procedure which will be helpful to understand
this work. \ In section III, we review an interesting application of
$S$-expansion procedure in order to get the Maxwell algebra family. \ Section
IV and V contain our main results. \ In section IV, we present the different
minimal $D=4$ Maxwell superalgebras which can be obtained from an
$S$-expansion of $\mathfrak{osp}\left(  4|1\right)  $ superalgebra. \ In
section V, we extend these results to the $N$-extended case and we present
different $N$-extended $D=4$ Maxwell superalgebras from an $S$-expansion of
the $\mathfrak{osp}\left(  4|N\right)  $ superalgebra. \ Section VI concludes
the work with a comment about possible developments.

\bigskip

\section{The $S$-expansion procedure}

\bigskip

\qquad In this section, we shall review the main aspects of the abelian
semigroup expansion method introduce in Ref. \cite{5}. \ Let us consider a Lie
(super)algebra $\mathfrak{g}$ with basis $T_{A}$ and a finite abelian
semigroup $S=\left\{  \lambda_{\alpha}\right\}  $. \ Then, the direct product
$\mathfrak{G}=S\times\mathfrak{g}$ is also a Lie algebra given by%
\begin{equation}
\left[  T_{\left(  A,\alpha\right)  },T_{\left(  B,\beta\right)  }\right]
=K_{\alpha\beta}^{\text{ \ }\gamma}C_{AB}^{\text{ \ \ }C}T_{\left(
C,\gamma\right)  }.
\end{equation}
The $S$-expansion procedure consist in combining the inner multiplication law
of a semigroup $S$ with the structure constants of a Lie algebra
$\mathfrak{g}$. \ Interestingly, there are differents ways of extracting
smaller algebras from $\mathfrak{G}=S\times\mathfrak{g}$. \ But before to
extract smaller algebras it is necessary to decompose the original algebra
$\mathfrak{g}$ into a direct sum of subspaces $\mathfrak{g}=\bigoplus
\nolimits_{p\in I}V_{p}$, where $I$ is a set of indices. \ Then for each
$p,q\in I$ it is possible to define $i_{\left(  p,q\right)  }\subset I$ such
that%
\begin{equation}
\left[  V_{p},V_{q}\right]  \subset\bigoplus\limits_{r\in i_{\left(
p,q\right)  }}V_{r}.
\end{equation}
Following the definitions of Ref. \cite{5}, it is possible to define a subset
decomposition $S=\bigcup\nolimits_{p\in I}S_{p}$ of the semigroup $S$ such
that%
\begin{equation}
S_{p}\cdot S_{q}\subset\bigcap\limits_{r\in i_{\left(  p,q\right)  }}S_{r}.
\end{equation}
When such subset decomposition exists, then we say that%
\begin{equation}
\mathfrak{G}_{R}=\bigoplus\limits_{p\in I}S_{p}\times V_{p},
\end{equation}
is a resonant subalgebra of $\mathfrak{G}=S\times\mathfrak{g}$.

Another case of smaller algebra is when there is a zero element in the
semigroup $0_{S}\in S$, such that for all $\lambda_{\alpha}\in S$, we have
$0_{S}\lambda_{\alpha}=0_{S}$. \ In this case, it is possible to reduce the
original algebra imposing $0_{S}\times T_{A}=0$ and obtain a new Lie (super)algebra.

Interestingly, there is a way to extract a reduced algebra from a resonant
subalgebra. \ Let $\mathfrak{G}_{R}=\bigoplus\nolimits_{p}S_{p}\times V_{p}$
be a resonant subalgebra of $\mathfrak{G}=S\times\mathfrak{g}$. \ Let
$S_{p}=\hat{S}_{p}\cup\check{S}_{p}$ be a partition of the subsets
$S_{p}\subset S$ such that%
\begin{align}
\hat{S}_{p}\cap\check{S}_{p}  &  =\varnothing,\\
\check{S}_{p}\cdot\hat{S}_{q}  &  \subset\bigcap\limits_{r\in i_{\left(
p,q\right)  }}\hat{S}_{r}.
\end{align}
Then, these conditions induce the decomposition%
\begin{align}
\mathfrak{\check{G}}_{R}  &  =\bigoplus\limits_{p\in I}\check{S}_{p}\times
V_{p},\\
\mathfrak{\hat{G}}_{R}  &  =\bigoplus\limits_{p\in I}\hat{S}_{p}\times V_{p},
\end{align}
with%
\begin{equation}
\left[  \mathfrak{\check{G}}_{R},\mathfrak{\hat{G}}_{R}\right]  \subset
\mathfrak{\hat{G}}_{R},
\end{equation}
and therefore $\left\vert \mathfrak{\check{G}}_{R}\right\vert $ corresponds to
a reduced algebra of $\mathfrak{G}_{R}$. \ The proofs of these definitions can
be found in Ref. \cite{5}.

\ We can see that if we want to obtain an $S$-expanded algebra, we only need
to solve the resonance condition for an abelian semigroup $S$. \ In the next
section we will briefly review an interesting application of $S$-expansion
procedure in order to derive the Maxwell algebra family. \ Then we shall see
how it is possible to extend this result to the supersymmetric case.

\bigskip

\section{Maxwell algebra as an $S$-expansion}

\bigskip

\qquad In order to describe how the $S$-expansion procedure works, let us
review here the results obtained in Refs. \cite{9, 10}. \ The symmetries of a
particle moving in a background in presence of a constant electromagnetic
field are described by the Maxwell algebra $\mathcal{M}$. \ This algebra is
provided by $\left\{  J_{ab},P_{a},Z_{ab}\right\}  $ where $\left\{
P_{a},J_{ab}\right\}  $ do not generate the Poincar\'{e} algebra. \ In fact a
particular characteristic of the Maxwell algebra is given by the relation%
\begin{equation}
\left[  P_{a},P_{b}\right]  =Z_{ab}%
\end{equation}
where $Z_{ab}$ commutes with all generators of the algebra except the Lorentz
generators $J_{ab}$,%
\begin{align}
\left[  J_{ab},Z_{cd}\right]   &  =\eta_{bc}J_{ad}-\eta_{ac}J_{bd}-\eta
_{bd}J_{ac}+\eta_{ad}J_{bc},\\
\left[  Z_{ab},P_{a}\right]   &  =\left[  Z_{ab},Z_{cd}\right]  =0.
\end{align}
The other commutators of the algebra are%
\begin{align}
\left[  J_{ab},J_{cd}\right]   &  =\eta_{bc}J_{ad}-\eta_{ac}J_{bd}-\eta
_{bd}J_{ac}+\eta_{ad}J_{bc},\\
\left[  J_{ab},P_{c}\right]   &  =\eta_{bc}P_{a}-\eta_{ac}P_{b}.
\end{align}

Following Refs. \cite{9,10}, it is possible to obtain the Maxwell algebra
$\mathcal{M}$ as an $S$-expansion of the $AdS$ Lie algebra $\mathfrak{g}$
using $S_{E}^{\left(  2\right)  }$ as the abelian semigroup.

Before to apply the $S$-expansion procedure it is necessary to consider a
decomposition of the original algebra $\mathfrak{g}$ in subspaces $V_{p}$,%
\begin{equation}
\mathfrak{g}=\mathfrak{so}\left(  3,2\right)  =\mathfrak{so}\left(
3,1\right)  \oplus\frac{\mathfrak{so}\left(  3,2\right)  }{\mathfrak{so}%
\left(  3,1\right)  }=V_{0}\oplus V_{1},
\end{equation}
where $V_{0}$ \ is generated by the Lorentz generator $\tilde{J}_{ab}$ and
$V_{1}$ is generated by the $AdS$ boost generator $\tilde{P}_{a}$. \ The
$\tilde{J}_{ab},$ $\tilde{P}_{a}$ generators satisfy the following relations%
\begin{align}
\left[  \tilde{J}_{ab},\tilde{J}_{cd}\right]   &  =\eta_{bc}\tilde{J}%
_{ad}-\eta_{ac}\tilde{J}_{bd}-\eta_{bd}\tilde{J}_{ac}+\eta_{ad}\tilde{J}%
_{bc},\label{ads01}\\
\left[  \tilde{J}_{ab},\tilde{P}_{c}\right]   &  =\eta_{bc}\tilde{P}_{a}%
-\eta_{ac}\tilde{P}_{b},\label{ads02}\\
\left[  \tilde{P}_{a},\tilde{P}_{b}\right]   &  =\tilde{J}_{ab}. \label{ads03}%
\end{align}
The subspace structure may be written as%
\begin{align}
\left[  V_{0},V_{0}\right]   &  \subset V_{0},\label{ec01}\\
\left[  V_{0},V_{1}\right]   &  \subset V_{1},\label{ec02}\\
\left[  V_{1},V_{1}\right]   &  \subset V_{0}. \label{ec03}%
\end{align}
Let $S_{E}^{\left(  2\right)  }=\left\{  \lambda_{0},\lambda_{1},\lambda
_{2},\lambda_{3}\right\}  $ be an abelian semigroup with the following subset
decomposition $S_{E}^{\left(  2\right)  }=S_{0}\cup S_{1}$, where the subsets
$S_{0},S_{1}$ are given by
\begin{align}
S_{0}  &  =\left\{  \lambda_{0},\lambda_{2},\lambda_{3}\right\}  ,\\
S_{1}  &  =\left\{  \lambda_{1},\lambda_{3}\right\}  ,
\end{align}
where $\lambda_{3}$ corresponds to the zero element of the semigroup $\left(
0_{s}=\lambda_{3}\right)  $. \ This subset decomposition is said to be
"resonant" because it satisfies [compare with eqs.$\left(  \ref{ec01}\right)
-\left(  \ref{ec03}\right)  $.]%
\begin{align}
S_{0}\cdot S_{0}  &  \subset S_{0},\\
S_{0}\cdot S_{1}  &  \subset S_{1},\\
S_{1}\cdot S_{1}  &  \subset S_{0.}%
\end{align}
In this case, the elements of the semigroup $\left\{  \lambda_{0},\lambda
_{1},\lambda_{2},\lambda_{3}\right\}  $ satisfy the following multiplication
law%
\begin{equation}
\lambda_{\alpha}\lambda_{\beta}=\left\{
\begin{array}
[c]{c}%
\lambda_{\alpha+\beta}\text{, \ \ \ \ when }\alpha+\beta\leq3,\\
\lambda_{3}\text{, \ \ \ \ \ \ when }\alpha+\beta>3.
\end{array}
\right.
\end{equation}

Following the definitions of Ref. \cite{5}, after extracting a resonant
subalgebra and performing its $0_{S}$-reduction, one finds the Maxwell algebra
$\mathcal{M}=\left\{  J_{ab},P_{a},Z_{ab}\right\}  $, whose generators can be
written in terms of the original ones,%
\begin{align}
J_{ab}  &  =\lambda_{0}\otimes\tilde{J}_{ab},\\
P_{a}  &  =\lambda_{1}\otimes\tilde{P}_{a},\\
Z_{ab}  &  =\lambda_{2}\otimes\tilde{J}_{ab.}%
\end{align}

Interestingly, in Refs. \cite{9, 11}, it was shown that standard
four-dimensional General Relativity emerges as a limit of a Born-Infeld theory
invariant under a certain subalgebra of the Maxwell algebra $\mathcal{M},$
which was denoted by%
\footnote{Initially called as $\mathcal{L}^{\mathfrak{B}_{4}}$ algebra.}
$\mathcal{L}^{\mathcal{M}}$. \ \ It was shown that this subalgebra can be
obtained as an $S$-expansion of the Lorentz algebra $\mathfrak{so}\left(
3,1\right)  $. \ \

It is possible to extend this procedure and obtain all the possible Maxwell
algebras type using the appropriate semigroup.

Following Ref. \cite{9}, let us consider the $S$-expansion of the Lie algebra
$\mathfrak{so}\left(  3,2\right)  $ using the abelian semigroup $S_{E}%
^{\left(  2n-1\right)  }=\left\{  \lambda_{0},\lambda_{1},\cdots,\lambda
_{2n}\right\}  $ with the following multiplication law%
\begin{equation}
\lambda_{\alpha}\lambda_{\beta}=\left\{
\begin{array}
[c]{c}%
\lambda_{\alpha+\beta}\text{, \ \ \ \ when }\alpha+\beta\leq2n,\\
\lambda_{2n}\text{, \ \ \ \ \ when }\alpha+\beta>2n.
\end{array}
\right.
\end{equation}

The $\lambda_{\alpha}$ elements are dimensionless and can be represented by
the set of $2n\times2n$ matrices $\left[  \lambda_{\alpha}\right]  _{\text{
\ }j}^{i}=\delta_{\text{ \ }j+\alpha}^{i}$, where $i,j=1,\cdots,2n-1$;
$\alpha=0,\cdots,2n$, and $\delta$ stands for the Kronecker delta.

After extracting a resonant subalgebra and performing its $0_{s}\left(
=\lambda_{2n}\right)  $-reduction, one find the Maxwell algebra type%
\footnote{Initially called as $\mathfrak{B}_{2n+1}$ algebra.}
$\mathcal{M}_{2n+1}$, whose generators are related to the original ones,%
\begin{align}
J_{ab}  &  =J_{\left(  ab,0\right)  }=\lambda_{0}\otimes\tilde{J}_{ab},\\
P_{a}  &  =P_{\left(  a,1\right)  }=\lambda_{1}\otimes\tilde{P}_{a},\\
Z_{ab}^{\left(  i\right)  }  &  =J_{\left(  ab,2i\right)  }=\lambda
_{2i}\otimes\tilde{J}_{ab},\\
Z_{a}^{\left(  i\right)  }  &  =P_{\left(  a,2i+1\right)  }=\lambda
_{2i+1}\otimes\tilde{P}_{a},
\end{align}
with $i=0,\cdots,n-1$. \ The commutators of the algebra are%
\begin{align}
\left[  P_{a},P_{b}\right]   &  =Z_{ab}^{\left(  1\right)  },\text{
\ \ \ \ }\left[  J_{ab},P_{c}\right]  =\eta_{bc}P_{a}-\eta_{ac}P_{b},\\
\left[  J_{ab,}J_{cd}\right]   &  =\eta_{cb}J_{ad}-\eta_{ca}J_{bd}+\eta
_{db}J_{ca}-\eta_{da}J_{cb},\\
\left[  J_{ab},Z_{c}^{\left(  i\right)  }\right]   &  =\eta_{bc}Z_{a}^{\left(
i\right)  }-\eta_{ac}Z_{b}^{\left(  i\right)  },\\
\left[  Z_{ab}^{\left(  i\right)  },P_{c}\right]   &  =\eta_{bc}Z_{a}^{\left(
i\right)  }-\eta_{ac}Z_{b}^{\left(  i\right)  },\\
\left[  Z_{ab}^{\left(  i\right)  },Z_{c}^{\left(  j\right)  }\right]   &
=\eta_{bc}Z_{a}^{\left(  i+j\right)  }-\eta_{ac}Z_{b}^{\left(  i+j\right)
},\\
\left[  J_{ab,}Z_{cd}^{\left(  i\right)  }\right]   &  =\eta_{cb}%
Z_{ad}^{\left(  i\right)  }-\eta_{ca}Z_{bd}^{\left(  i\right)  }+\eta
_{db}Z_{ca}^{\left(  i\right)  }-\eta_{da}Z_{cb}^{\left(  i\right)  },\\
\left[  Z_{ab,}^{\left(  i\right)  }Z_{cd}^{\left(  j\right)  }\right]   &
=\eta_{cb}Z_{ad}^{\left(  i+j\right)  }-\eta_{ca}Z_{bd}^{\left(  i+j\right)
}+\eta_{db}Z_{ca}^{\left(  i+j\right)  }-\eta_{da}Z_{cb}^{\left(  i+j\right)
},\\
\left[  P_{a},Z_{c}^{\left(  i\right)  }\right]   &  =Z_{ab}^{\left(
i+1\right)  },\text{ \ \ \ \ }\left[  Z_{a}^{\left(  i\right)  }%
,Z_{c}^{\left(  j\right)  }\right]  =Z_{ab}^{\left(  i+j+1\right)  }.
\end{align}

We note that setting $Z_{ab}^{\left(  i+1\right)  }$ and $Z_{a}^{\left(
i\right)  }$ equal to zero, we reobtain the Maxwell algebra $\mathcal{M}$.
\ In fact, every Maxwell algebra type $\mathcal{M}_{l}$ can be obtained from
$\mathcal{M}_{2m+1}$ setting some bosonic generators equal to zero. \ These
algebras are particularly interesting in gravity context, since it was shown
in \cite{9} that standard odd-dimensional general relativity may emerge as the
weak coupling constant limit of $\left(  2p+1\right)  $-dimensional
Chern-Simons Lagrangian invariant under the Maxwell algebra type
$\mathcal{M}_{2m+1}$, if and only if $m\geq p$. \ Similarly, it was shown that
standard even-dimensional general relativity emerges as the weak coupling
constant limit of a $\left(  2p\right)  $-dimensional Born-Infeld type
Lagrangian invariant under a subalgebra $\mathcal{L}^{\mathcal{M}_{2m}}$ of
the Maxwell algebra type, if and only if $m\geq p$.

\bigskip

\section{S-expansion of the $\mathfrak{osp}\left(  4|1\right)  $ superalgebra}

\bigskip

\qquad In this section, we shall take the $AdS$ superalgebra $\mathfrak{osp}%
\left(  4|1\right)  $ as a starting point. \ We will see that different
choices of abelian semigroup $S$ lead to new interesting $D=4$ superalgebras.
\ In every case, before to apply the $S$-expansion procedure it is necessary
to decompose the original algebra $\mathfrak{g}$ as a direct sum of subspaces
$V_{p}$,%
\begin{align}
\mathfrak{g}=\mathfrak{osp}\left(  4|1\right)   &  =\mathfrak{so}\left(
3,1\right)  \oplus\frac{\mathfrak{osp}\left(  4|1\right)  }{\mathfrak{sp}%
\left(  4\right)  }\oplus\frac{\mathfrak{sp}\left(  4\right)  }{\mathfrak{so}%
\left(  3,1\right)  }\nonumber\\
&  =V_{0}\oplus V_{1}\oplus V_{2},
\end{align}
where $V_{0}$ corresponds to the Lorentz subspace generated by $\tilde{J}%
_{ab}$, $V_{1}$ corresponds to the fermionic subspace generated by$\ $a
$4$-component Majorana spinor charge $\tilde{Q}_{\alpha}$ and $V_{2}$
corresponds to the $AdS$ boost generated by $\tilde{P}_{a}$. \ The
$\mathfrak{osp}\left(  4|1\right)  $ (anti)commutation relations read%
\begin{align}
\left[  \tilde{J}_{ab},\tilde{J}_{cd}\right]   &  =\eta_{bc}\tilde{J}%
_{ad}-\eta_{ac}\tilde{J}_{bd}-\eta_{bd}\tilde{J}_{ac}+\eta_{ad}\tilde{J}%
_{bc},\label{ads1}\\
\left[  \tilde{J}_{ab},\tilde{P}_{c}\right]   &  =\eta_{bc}\tilde{P}_{a}%
-\eta_{ac}\tilde{P}_{b},\\
\left[  \tilde{P}_{a},\tilde{P}_{b}\right]   &  =\tilde{J}_{ab},\\
\left[  \tilde{J}_{ab},\tilde{Q}_{\alpha}\right]   &  =-\frac{1}{2}\left(
\gamma_{ab}\tilde{Q}\right)  _{\alpha},\text{ \ \ \ \ }\left[  \tilde{P}%
_{a},\tilde{Q}_{\alpha}\right]  =-\frac{1}{2}\left(  \gamma_{a}\tilde
{Q}\right)  _{\alpha},\\
\left\{  \tilde{Q}_{\alpha},\tilde{Q}_{\beta}\right\}   &  =-\frac{1}%
{2}\left[  \left(  \gamma^{ab}C\right)  _{\alpha\beta}\tilde{J}_{ab}-2\left(
\gamma^{a}C\right)  _{\alpha\beta}\tilde{P}_{a}\right]  . \label{ads5}%
\end{align}
Here, $C$ stands for the charge conjugation matrix and $\gamma_{a}$ are Dirac matrices.

The subspace structure may be written as%
\begin{align}
\left[  V_{0},V_{0}\right]   &  \subset V_{0},\label{ec04}\\
\left[  V_{0},V_{1}\right]   &  \subset V_{1},\label{ec05}\\
\left[  V_{0},V_{2}\right]   &  \subset V_{2},\label{ec06}\\
\left[  V_{1},V_{1}\right]   &  \subset V_{0}\oplus V_{2},\label{ec07}\\
\left[  V_{1},V_{2}\right]   &  \subset V_{1,}\label{ec08}\\
\left[  V_{2},V_{2}\right]   &  \subset V_{0}. \label{ec09}%
\end{align}

The next step consists of finding a subset decomposition of a semigroup $S$
which is "resonant" with respect to $\left(  \ref{ec04}\right)  -\left(
\ref{ec09}\right)  $.

\bigskip

\subsection{Minimal $D=4$ superMaxwell algebra}

\bigskip

\qquad Let us consider $S_{E}^{\left(  4\right)  }=\left\{  \lambda
_{0},\lambda_{1},\lambda_{2},\lambda_{3},\lambda_{4},\lambda_{5}\right\}  $ as
the relevant finite abelian semigroup whose elements\ are dimensionless and
obey the multiplication law%
\begin{equation}
\lambda_{\alpha}\lambda_{\beta}=\left\{
\begin{array}
[c]{c}%
\lambda_{\alpha+\beta}\text{, \ \ \ \ when }\alpha+\beta\leq5,\\
\lambda_{5}\text{, \ \ \ \ \ \ \ when }\alpha+\beta>5.
\end{array}
\right.  \label{lm01}%
\end{equation}
In this case, $\lambda_{5}$ plays the role of the zero element of the
semigroup $S_{E}^{\left(  4\right)  }$, so we have for each $\lambda_{\alpha
}\in S_{E}^{\left(  4\right)  },$ $\lambda_{5}\lambda_{\alpha}=\lambda
_{5}=0_{s}$. \ Let us consider the decomposition $S=S_{0}\cup S_{1}\cup
S_{2},$ with%
\begin{align}
S_{0}  &  =\left\{  \lambda_{0},\lambda_{2},\lambda_{4},\lambda_{5}\right\}
,\\
S_{1}  &  =\left\{  \lambda_{1},\lambda_{3},\lambda_{5}\right\}  ,\\
S_{2}  &  =\left\{  \lambda_{2},\lambda_{4},\lambda_{5}\right\}  .
\end{align}
One sees that this decomposition is resonant since it satisfies the same
structure as the subspaces $V_{p}$ [compare with eqs. $\left(  \ref{ec04}%
\right)  -\left(  \ref{ec09}\right)  $]%
\begin{align}
S_{0}\cdot S_{0}  &  \subset S_{0},\label{D01}\\
S_{0}\cdot S_{1}  &  \subset S_{1},\\
S_{0}\cdot S_{2}  &  \subset S_{2},\\
S_{1}\cdot S_{1}  &  \subset S_{0}\cap S_{2},\\
S_{1}\cdot S_{2}  &  \subset S_{1},\\
S_{2}\cdot S_{2}  &  \subset S_{0}. \label{D06}%
\end{align}

Following theorem IV.2 of Ref. \cite{5}, we can say that the superalgebra
\begin{equation}
\mathfrak{G}_{R}=W_{0}\oplus W_{1}\oplus W_{2}\text{,}%
\end{equation}
is a resonant super subalgebra of $S_{E}^{\left(  4\right)  }\times
\mathfrak{g}$, where%
\begin{align}
W_{0}  &  =\left(  S_{0}\times V_{0}\right)  =\left\{  \lambda_{0},\lambda
_{2},\lambda_{4},\lambda_{5}\right\}  \times\left\{  \tilde{J}_{ab}\right\}
=\left\{  \lambda_{0}\tilde{J}_{ab},\lambda_{2}\tilde{J}_{ab},\lambda
_{4}\tilde{J}_{ab},\lambda_{5}\tilde{J}_{ab}\right\}  ,\\
W_{1}  &  =\left(  S_{1}\times V_{1}\right)  =\left\{  \lambda_{1},\lambda
_{3},\lambda_{5}\right\}  \times\left\{  \tilde{Q}_{\alpha}\right\}  =\left\{
\lambda_{1}\tilde{Q}_{\alpha},\lambda_{3}\tilde{Q}_{\alpha},\lambda_{5}%
\tilde{Q}_{\alpha}\right\}  ,\\
W_{2}  &  =\left(  S_{2}\times V_{2}\right)  =\left\{  \lambda_{2},\lambda
_{4},\lambda_{5}\right\}  \times\left\{  \tilde{P}_{a}\right\}  =\left\{
\lambda_{2}\tilde{P}_{a},\lambda_{4}\tilde{P}_{a},\lambda_{5}\tilde{P}%
_{a}\right\}  .
\end{align}

In order to extract a smaller superalgebra from the resonant super subalgebra
$\mathfrak{G}_{R}$ it is necessary to apply the reduction procedure.

Let $S_{p}=\hat{S}_{p}\cup\check{S}_{p}$ be a partition of the subsets
$S_{p}\subset S$ where%
\begin{align}
\check{S}_{0}  &  =\left\{  \lambda_{0},\lambda_{2},\lambda_{4}\right\}
,\text{ \ \ \ \ }\hat{S}_{0}=\left\{  \lambda_{5}\right\}  ,\\
\check{S}_{1}  &  =\left\{  \lambda_{1},\lambda_{3}\right\}  ,\text{
\ \ \ \ }\hat{S}_{1}=\left\{  \lambda_{5}\right\}  ,\\
\check{S}_{2}  &  =\left\{  \lambda_{2}\right\}  ,\text{ \ \ \ \ \ \ \ \ }%
\hat{S}_{2}=\left\{  \lambda_{4},\lambda_{5}\right\}  .
\end{align}
For each $p$, $\hat{S}_{p}\cap\check{S}_{p}=\varnothing$, and using the
product $\left(  \ref{lm01}\right)  $ one sees that the partition satisfies
[compare with ecs. $\left(  \ref{ec04}\right)  -\left(  \ref{ec09}\right)  $]%
\begin{equation}%
\begin{tabular}
[c]{ll}%
$\check{S}_{0}\cdot\hat{S}_{0}\subset\hat{S}_{0},$ & $\check{S}_{1}\cdot
\hat{S}_{1}\subset\hat{S}_{0}\cap\hat{S}_{2},$\\
$\check{S}_{0}\cdot\hat{S}_{1}\subset\hat{S}_{1},$ & $\check{S}_{1}\cdot
\hat{S}_{2}\subset\hat{S}_{1},$\\
$\check{S}_{0}\cdot\hat{S}_{2}\subset\hat{S}_{2},$ & $\check{S}_{2}\cdot
\hat{S}_{2}\subset\hat{S}_{0}.$%
\end{tabular}
\end{equation}
Then, following definitions of Ref. \cite{5}, we have%
\begin{align}
\mathfrak{\check{G}}_{R}  &  =\left(  \check{S}_{0}\times V_{0}\right)
\oplus\left(  \check{S}_{1}\times V_{1}\right)  \oplus\left(  \check{S}%
_{2}\times V_{2}\right)  ,\\
\mathfrak{\hat{G}}_{R}  &  =\left(  \hat{S}_{0}\times V_{0}\right)
\oplus\left(  \hat{S}_{1}\times V_{1}\right)  \oplus\left(  \hat{S}_{2}\times
V_{2}\right)  ,
\end{align}
where%
\begin{equation}
\left[  \mathfrak{\check{G}}_{R},\mathfrak{\hat{G}}_{R}\right]  \subset
\mathfrak{\hat{G}}_{R},
\end{equation}
and therefore $\left\vert \mathfrak{\check{G}}_{R}\right\vert $ corresponds to
a reduced algebra of $\mathfrak{G}_{R}$. These $S$-expansion process can be
seen explicitly in the following diagrams:%
\begin{equation}%
\begin{tabular}
[c]{rrrr}\cline{2-4}\cline{4-4}%
$\lambda_{5}$ & \multicolumn{1}{|r}{$J_{ab,5}$} &
\multicolumn{1}{|r}{$Q_{\alpha,5}$} & \multicolumn{1}{|r|}{$P_{a,5}$%
}\\\cline{2-4}\cline{4-4}%
$\lambda_{4}$ & \multicolumn{1}{|r}{$J_{ab,4}$} & \multicolumn{1}{|r}{} &
\multicolumn{1}{|r|}{$P_{a,4}$}\\\cline{2-4}\cline{4-4}%
$\lambda_{3}$ & \multicolumn{1}{|r}{} & \multicolumn{1}{|r}{$Q_{\alpha,3}$} &
\multicolumn{1}{|r|}{}\\\cline{2-4}\cline{4-4}%
$\lambda_{2}$ & \multicolumn{1}{|r}{$J_{ab,2}$} & \multicolumn{1}{|r}{} &
\multicolumn{1}{|r|}{$P_{a,2}$}\\\cline{2-4}\cline{4-4}%
$\lambda_{1}$ & \multicolumn{1}{|r}{} & \multicolumn{1}{|r}{$Q_{\alpha,1}$} &
\multicolumn{1}{|r|}{}\\\cline{2-4}\cline{4-4}%
$\lambda_{0}$ & \multicolumn{1}{|r}{$J_{ab,0}$} & \multicolumn{1}{|r}{} &
\multicolumn{1}{|r|}{}\\\cline{2-4}\cline{4-4}
& $V_{0}$ & $V_{1}$ & $V_{2}$%
\end{tabular}
\ \ \ \ \text{ \ \ \ \ \ \ \ \ \ \ \ \ \ \ }%
\begin{tabular}
[c]{rrrr}\cline{2-4}\cline{4-4}%
$\lambda_{5}$ & \multicolumn{1}{|r}{} & \multicolumn{1}{|r}{} &
\multicolumn{1}{|r|}{}\\\cline{2-4}\cline{4-4}%
$\lambda_{4}$ & \multicolumn{1}{|r}{$J_{ab,4}$} & \multicolumn{1}{|r}{} &
\multicolumn{1}{|r|}{}\\\cline{2-4}\cline{4-4}%
$\lambda_{3}$ & \multicolumn{1}{|r}{} & \multicolumn{1}{|r}{$Q_{\alpha,3}$} &
\multicolumn{1}{|r|}{}\\\cline{2-4}\cline{4-4}%
$\lambda_{2}$ & \multicolumn{1}{|r}{$J_{ab,2}$} & \multicolumn{1}{|r}{} &
\multicolumn{1}{|r|}{$P_{a,2}$}\\\cline{2-4}\cline{4-4}%
$\lambda_{1}$ & \multicolumn{1}{|r}{} & \multicolumn{1}{|r}{$Q_{\alpha,1}$} &
\multicolumn{1}{|r|}{}\\\cline{2-4}\cline{4-4}%
$\lambda_{0}$ & \multicolumn{1}{|r}{$J_{ab,0}$} & \multicolumn{1}{|r}{} &
\multicolumn{1}{|r|}{}\\\cline{2-4}\cline{4-4}
& $V_{0}$ & $V_{1}$ & $V_{2}$%
\end{tabular}
\ \ \ \ ,
\end{equation}
where we have defined $J_{ab,i}=\lambda_{i}\tilde{J}_{ab},$ $P_{a,i}%
=\lambda_{i}\tilde{P}_{a}$ and $Q_{\alpha,i}=\lambda_{i}\tilde{Q}_{\alpha}$.
\ We can observe that the first diagram corresponds to the resonant subalgebra
of the $S$-expanded superalgebra $S_{E}^{\left(  4\right)  }\times$
$\mathfrak{osp}\left(  4|1\right)  $. \ The second one consists in a
particular reduction of the resonant subalgebra.

Thus, the new superalgebra obtained is generated by $\left\{  J_{ab}%
,P_{a},\tilde{Z}_{ab},Z_{ab},Q_{\alpha},\Sigma_{\alpha}\right\}  $ where these
new generators can be written as%
\begin{equation}%
\begin{tabular}
[c]{ll}%
$J_{ab}=J_{ab,0}=\lambda_{0}\tilde{J}_{ab},$ & $P_{a}=P_{a,2}=\lambda
_{2}\tilde{P}_{a},$\\
$\tilde{Z}_{ab}=J_{ab,2}=\lambda_{2}\tilde{J}_{ab},$ & $Z_{ab}=J_{ab,4}%
=\lambda_{4}\tilde{J}_{ab},$\\
$Q_{\alpha}=Q_{\alpha,1}=\lambda_{1}\tilde{Q}_{\alpha},$ & $\Sigma_{\alpha
}=Q_{\alpha,3}=\lambda_{3}\tilde{Q}_{\alpha}.$%
\end{tabular}
\end{equation}
These new generators satisfy the commutation relations%
\begin{align}
\left[  J_{ab},J_{cd}\right]   &  =\eta_{bc}J_{ad}-\eta_{ac}J_{bd}-\eta
_{bd}J_{ac}+\eta_{ad}J_{bc},\\
\left[  J_{ab},P_{c}\right]   &  =\eta_{bc}P_{a}-\eta_{ac}P_{b},\text{
\ \ \ \ }\left[  P_{a},P_{b}\right]  =Z_{ab},\\
\left[  J_{ab},Z_{cd}\right]   &  =\eta_{bc}Z_{ad}-\eta_{ac}Z_{bd}-\eta
_{bd}Z_{ac}+\eta_{ad}Z_{bc},\\
\left[  P_{a},Q_{\alpha}\right]   &  =-\frac{1}{2}\left(  \gamma_{a}%
\Sigma\right)  _{\alpha},\\
\left[  J_{ab},Q_{\alpha}\right]   &  =-\frac{1}{2}\left(  \gamma
_{ab}Q\right)  _{\alpha},\\
\left[  J_{ab},\Sigma_{\alpha}\right]   &  =-\frac{1}{2}\left(  \gamma
_{ab}\Sigma\right)  _{\alpha},\\
\left\{  Q_{\alpha},Q_{\beta}\right\}   &  =-\frac{1}{2}\left[  \left(
\gamma^{ab}C\right)  _{\alpha\beta}\tilde{Z}_{ab}-2\left(  \gamma^{a}C\right)
_{\alpha\beta}P_{a}\right]  ,\\
\left\{  Q_{\alpha},\Sigma_{\beta}\right\}   &  =-\frac{1}{2}\left(
\gamma^{ab}C\right)  _{\alpha\beta}Z_{ab},
\end{align}%
\begin{align}
\left[  J_{ab},\tilde{Z}_{ab}\right]   &  =\eta_{bc}\tilde{Z}_{ad}-\eta
_{ac}\tilde{Z}_{bd}-\eta_{bd}\tilde{Z}_{ac}+\eta_{ad}\tilde{Z}_{bc},\\
\left[  \tilde{Z}_{ab},\tilde{Z}_{cd}\right]   &  =\eta_{bc}Z_{ad}-\eta
_{ac}Z_{bd}-\eta_{bd}Z_{ac}+\eta_{ad}Z_{bc},\\
\left[  \tilde{Z}_{ab},Q_{\alpha}\right]   &  =-\frac{1}{2}\left(  \gamma
_{ab}\Sigma\right)  _{\alpha},\\
\text{others}  &  =0,
\end{align}
where we have used the multiplication law of the semigroup $\left(
\ref{lm01}\right)  $ and the commutation relations of the original
superalgebra (see Appendix A). \ The new superalgebra obtained after a reduced
resonant $S$-expansion of $\mathfrak{osp}\left(  4|1\right)  $ superalgebra
corresponds to a generalized minimal superMaxwell algebra $s\mathcal{M}_{4}$
in $D=4$ . \ One can see that imposing $\tilde{Z}_{ab}=0$ leads us to the
minimal superMaxwell algebra $s\mathcal{M}$ \cite{17, 19}. \ This can be done
since the Jacobi identities for spinors generators are satisfied due to the
gamma matrix identity $\left(  C\gamma^{a}\right)  _{\left(  \alpha
\beta\right.  }\left(  C\gamma_{a}\right)  _{\left.  \gamma\delta\right)  }=0$
$\left(  \text{cyclic permutations of }\alpha,\beta,\gamma\right)  $.

In this case, the $S$-expansion procedure produces a new Majorana spinor
charge $\Sigma$. \ The introduction of a second abelian spinorial generator
has been initially proposed in Ref. \cite{7} in the context of $D=11$
supergravity and subsequently in Ref. \cite{22} in the context of superstring
theory. \

The \ $s\mathcal{M}$ superalgebra seems particularly interesting in the
context of $D=4$ $\ $supergravity. In fact in Ref. \cite{21}, it was shown
that $D=4$, $N=1$ pure supergravity lagrangian can be written as a quadratic
expression in the curvatures of the gauge fields associated with the minimal
superMaxwell algebra.

It is interesting to note that the expanded superalgebra contains the Maxwell
algebra $\mathcal{M}=\left\{  J_{ab},P_{a},Z_{ab}\right\}  $ and the Lorentz
type subalgebra $\mathcal{L}^{\mathcal{M}}=\left\{  J_{ab},Z_{ab}\right\}  $
introduced in Ref. \cite{11} as subalgebras.

\bigskip

\subsection{ Minimal $D=4$ superMaxwell algebra type $s\mathcal{M}_{5}$}

\bigskip

\qquad In Ref. \cite{9} , it was shown that the Maxwell algebra type
$\mathcal{M}_{m}$ can be obtained from an $S$-expansion of $AdS$ algebra.
\ These bigger algebras require semigroups with more elements but with the
same type of multiplication law. \ Since our main motivation is to obtain a
$D=4$ superMaxwell algebra type $s\mathcal{M}_{m}$ it seems natural to
consider a semigroup bigger than $S_{E}^{\left(  4\right)  }=\left\{
\lambda_{0},\lambda_{1},\lambda_{2},\lambda_{3},\lambda_{4},\lambda
_{5}\right\}  $. \ \ As in the previous case, we shall consider $\mathfrak{g}%
=\mathfrak{osp}\left(  4|1\right)  $ as a starting point with the subspace
structure given by eqs. $\left(  \ref{ec04}\right)  -\left(  \ref{ec09}%
\right)  $.

Let us consider $S_{E}^{\left(  6\right)  }=\left\{  \lambda_{0},\lambda
_{1},\lambda_{2},\lambda_{3},\lambda_{4},\lambda_{5},\lambda_{6},\lambda
_{7}\right\}  $ as the relevant finite abelian semigroup whose elements are
dimensionless and obey the multiplication law%
\begin{equation}
\lambda_{\alpha}\lambda_{\beta}=\left\{
\begin{array}
[c]{c}%
\lambda_{\alpha+\beta}\text{, \ \ \ \ when }\alpha+\beta\leq7,\\
\lambda_{7}\text{, \ \ \ \ \ \ \ when }\alpha+\beta>7,
\end{array}
\right.  \label{lm02}%
\end{equation}
where $\lambda_{7}$ plays the role of the zero element of the semigroup
$S_{E}^{\left(  6\right)  }$. \ \ Let us consider the decomposition
$S=S_{0}\cup S_{1}\cup S_{2},$ with%
\begin{align}
S_{0}  &  =\left\{  \lambda_{0},\lambda_{2},\lambda_{4},\lambda_{6}%
,\lambda_{7}\right\}  ,\\
S_{1}  &  =\left\{  \lambda_{1},\lambda_{3},\lambda_{5},\lambda_{7}\right\}
,\\
S_{2}  &  =\left\{  \lambda_{2},\lambda_{4},\lambda_{6},\lambda_{7}\right\}  .
\end{align}
This subset decomposition of $S_{E}^{\left(  6\right)  }$ satisfies the
resonance condition since it satisfies the same structure that the subspaces
$V_{p}$ [compare with eqs. $\left(  \ref{ec04}\right)  -\left(  \ref{ec09}%
\right)  $]%
\begin{align}
S_{0}\cdot S_{0}\subset S_{0},\text{ \ \ \ \ \ \ \ }  &  S_{1}\cdot
S_{1}\subset S_{0}\cap S_{2},\\
S_{0}\cdot S_{1}\subset S_{1},\text{ \ \ \ \ \ \ \ }  &  S_{1}\cdot
S_{2}\subset S_{1},\\
S_{0}\cdot S_{2}\subset S_{2},\text{ \ \ \ \ \ \ \ }  &  S_{2}\cdot
S_{2}\subset S_{0}.
\end{align}

Therefore, according to Theorem IV.2 of Ref. \cite{5}, we have that%
\begin{equation}
\mathfrak{G}_{R}=W_{0}+W_{1}+W_{2},
\end{equation}
with%
\begin{equation}
W_{p}=S_{p}\times V_{p},
\end{equation}
is a resonant super subalgebra of $\mathfrak{G}=S\times\mathfrak{g}$.

As in the previous case, it is possible to extract a smaller superalgebra from
the resonant super subalgebra $\mathfrak{G}_{R}$ using the reduction
procedure. \ Let $S_{p}=\hat{S}_{p}\cup\check{S}_{p}$ be a partition of the
subsets $S_{p}\subset S$ where%
\begin{align}
\check{S}_{0}  &  =\left\{  \lambda_{0},\lambda_{2},\lambda_{4}\right\}
,\text{ \ \ \ \ }\hat{S}_{0}=\left\{  \lambda_{6},\lambda_{7}\right\}  ,\\
\check{S}_{1}  &  =\left\{  \lambda_{1},\lambda_{3},\lambda_{5}\right\}
,\text{ \ \ \ \ }\hat{S}_{1}=\left\{  \lambda_{7}\right\}  ,\\
\check{S}_{2}  &  =\left\{  \lambda_{2},\lambda_{4},\lambda_{6}\right\}
,\text{ \ \ \ \ \ \ \ \ }\hat{S}_{2}=\left\{  \lambda_{7}\right\}  .
\end{align}
For each $p$, $\hat{S}_{p}\cap\check{S}_{p}=\varnothing$, and using the
product $\left(  \ref{lm02}\right)  $ one sees that the partition satisfies
[compare with ecs. $\left(  \ref{ec04}\right)  -\left(  \ref{ec09}\right)  $]%
\begin{equation}%
\begin{tabular}
[c]{ll}%
$\check{S}_{0}\cdot\hat{S}_{0}\subset\hat{S}_{0},$ & $\check{S}_{1}\cdot
\hat{S}_{1}\subset\hat{S}_{0}\cap\hat{S}_{2},$\\
$\check{S}_{0}\cdot\hat{S}_{1}\subset\hat{S}_{1},$ & $\check{S}_{1}\cdot
\hat{S}_{2}\subset\hat{S}_{1},$\\
$\check{S}_{0}\cdot\hat{S}_{2}\subset\hat{S}_{2},$ & $\check{S}_{2}\cdot
\hat{S}_{2}\subset\hat{S}_{0}.$%
\end{tabular}
\end{equation}
Then, we have%
\begin{align}
\mathfrak{\check{G}}_{R}  &  =\left(  \check{S}_{0}\times V_{0}\right)
\oplus\left(  \check{S}_{1}\times V_{1}\right)  \oplus\left(  \check{S}%
_{2}\times V_{2}\right)  ,\\
\mathfrak{\hat{G}}_{R}  &  =\left(  \hat{S}_{0}\times V_{0}\right)
\oplus\left(  \hat{S}_{1}\times V_{1}\right)  \oplus\left(  \hat{S}_{2}\times
V_{2}\right)  ,
\end{align}
where%
\begin{equation}
\left[  \mathfrak{\check{G}}_{R},\mathfrak{\hat{G}}_{R}\right]  \subset
\mathfrak{\hat{G}}_{R},
\end{equation}
and therefore $\left\vert \mathfrak{\check{G}}_{R}\right\vert $ corresponds to
a reduced algebra of $\mathfrak{G}_{R}$. These procedures can be saw
explicitly in the following diagrams:%
\begin{equation}%
\begin{tabular}
[c]{rrrr}\cline{2-4}\cline{4-4}%
$\lambda_{7}$ & \multicolumn{1}{|r}{$J_{ab,7}$} &
\multicolumn{1}{|r}{$Q_{\alpha,7}$} & \multicolumn{1}{|r|}{$P_{a,7}$%
}\\\cline{2-4}%
$\lambda_{6}$ & \multicolumn{1}{|r}{$J_{ab,6}$} & \multicolumn{1}{|r}{} &
\multicolumn{1}{|r|}{$P_{a,6}$}\\\cline{2-4}%
$\lambda_{5}$ & \multicolumn{1}{|r}{} & \multicolumn{1}{|r}{$Q_{\alpha,5}$} &
\multicolumn{1}{|r|}{}\\\cline{2-4}%
$\lambda_{4}$ & \multicolumn{1}{|r}{$J_{ab,4}$} & \multicolumn{1}{|r}{} &
\multicolumn{1}{|r|}{$P_{a,4}$}\\\cline{2-4}\cline{4-4}%
$\lambda_{3}$ & \multicolumn{1}{|r}{} & \multicolumn{1}{|r}{$Q_{\alpha,3}$} &
\multicolumn{1}{|r|}{}\\\cline{2-4}\cline{4-4}%
$\lambda_{2}$ & \multicolumn{1}{|r}{$J_{ab,2}$} & \multicolumn{1}{|r}{} &
\multicolumn{1}{|r|}{$P_{a,2}$}\\\cline{2-4}\cline{4-4}%
$\lambda_{1}$ & \multicolumn{1}{|r}{} & \multicolumn{1}{|r}{$Q_{\alpha,1}$} &
\multicolumn{1}{|r|}{}\\\cline{2-4}\cline{4-4}%
$\lambda_{0}$ & \multicolumn{1}{|r}{$J_{ab,0}$} & \multicolumn{1}{|r}{} &
\multicolumn{1}{|r|}{}\\\cline{2-4}\cline{4-4}
& $V_{0}$ & $V_{1}$ & $V_{2}$%
\end{tabular}
\ \ \ \text{ \ \ \ \ \ \ \ \ \ \ \ \ \ \ }%
\begin{tabular}
[c]{rrrr}\cline{2-4}\cline{4-4}%
$\lambda_{7}$ & \multicolumn{1}{|r}{} & \multicolumn{1}{|r}{} &
\multicolumn{1}{|r|}{}\\\cline{2-4}%
$\lambda_{6}$ & \multicolumn{1}{|r}{} & \multicolumn{1}{|r}{} &
\multicolumn{1}{|r|}{$P_{a,6}$}\\\cline{2-4}%
$\lambda_{5}$ & \multicolumn{1}{|r}{} & \multicolumn{1}{|r}{$Q_{\alpha,5}$} &
\multicolumn{1}{|r|}{}\\\cline{2-4}%
$\lambda_{4}$ & \multicolumn{1}{|r}{$J_{ab,4}$} & \multicolumn{1}{|r}{} &
\multicolumn{1}{|r|}{$P_{a,4}$}\\\cline{2-4}\cline{4-4}%
$\lambda_{3}$ & \multicolumn{1}{|r}{} & \multicolumn{1}{|r}{$Q_{\alpha,3}$} &
\multicolumn{1}{|r|}{}\\\cline{2-4}\cline{4-4}%
$\lambda_{2}$ & \multicolumn{1}{|r}{$J_{ab,2}$} & \multicolumn{1}{|r}{} &
\multicolumn{1}{|r|}{$P_{a,2}$}\\\cline{2-4}\cline{4-4}%
$\lambda_{1}$ & \multicolumn{1}{|r}{} & \multicolumn{1}{|r}{$Q_{\alpha,1}$} &
\multicolumn{1}{|r|}{}\\\cline{2-4}\cline{4-4}%
$\lambda_{0}$ & \multicolumn{1}{|r}{$J_{ab,0}$} & \multicolumn{1}{|r}{} &
\multicolumn{1}{|r|}{}\\\cline{2-4}\cline{4-4}
& $V_{0}$ & $V_{1}$ & $V_{2}$%
\end{tabular}
\ \ \ ,
\end{equation}
where we have defined $J_{ab,i}=\lambda_{i}\tilde{J}_{ab},$ $P_{a,i}%
=\lambda_{i}\tilde{P}_{a}$ and $Q_{\alpha,i}=\lambda_{i}\tilde{Q}_{\alpha}$.
\ The first diagram corresponds to the resonant subalgebra of the $S$-expanded
superalgebra $S_{E}^{\left(  6\right)  }\times$ $\mathfrak{osp}\left(
4|1\right)  $. \ The second one consists in a particular reduction of the
resonant subalgebra.

The new superalgebra is generated by $\left\{  J_{ab},P_{a},Z_{ab},\tilde
{Z}_{ab},Z_{a},\tilde{Z}_{a},Q_{\alpha},\Sigma_{\alpha},\Phi_{\alpha}\right\}
$ where these new generators can be written as%
\begin{equation}%
\begin{tabular}
[c]{ll}%
$J_{ab}=J_{ab,0}=\lambda_{0}\tilde{J}_{ab},$ & $\tilde{Z}_{a}=P_{a,4}%
=\lambda_{4}\tilde{P}_{a},$\\
$P_{a}=P_{a,2}=\lambda_{2}\tilde{P}_{a},$ & $Q_{\alpha}=Q_{\alpha,1}%
=\lambda_{1}\tilde{Q}_{\alpha},$\\
$Z_{ab}=J_{ab,4}=\lambda_{4}\tilde{J}_{ab},$ & $\Sigma_{\alpha}=Q_{\alpha
,3}=\lambda_{3}\tilde{Q}_{\alpha},$\\
$\tilde{Z}_{ab}=J_{ab,2}=\lambda_{2}\tilde{J}_{ab},$ & $\Phi_{\alpha
}=Q_{\alpha,5}=\lambda_{5}\tilde{Q}_{\alpha},$\\
$Z_{a}=P_{a,6}=\lambda_{6}\tilde{P}_{a}.$ &
\end{tabular}
\end{equation}
These new generators satisfy the commutation relations%
\begin{align}
\left[  J_{ab},J_{cd}\right]   &  =\eta_{bc}J_{ad}-\eta_{ac}J_{bd}-\eta
_{bd}J_{ac}+\eta_{ad}J_{bc},\label{SM501}\\
\left[  J_{ab},P_{c}\right]   &  =\eta_{bc}P_{a}-\eta_{ac}P_{b},\text{
\ \ \ \ \ \ }\left[  P_{a},P_{b}\right]  =Z_{ab},\\
\left[  J_{ab},Z_{cd}\right]   &  =\eta_{bc}Z_{ad}-\eta_{ac}Z_{bd}-\eta
_{bd}Z_{ac}+\eta_{ad}Z_{bc},\\
\left[  Z_{ab},P_{c}\right]   &  =\eta_{bc}Z_{a}-\eta_{ac}Z_{b},\text{
\ \ \ \ \ \ }\left[  J_{ab},Z_{c}\right]  =\eta_{bc}Z_{a}-\eta_{ac}%
Z_{b},\label{SM504}\\
\left[  \tilde{Z}_{ab},\tilde{Z}_{cd}\right]   &  =\eta_{bc}Z_{ad}-\eta
_{ac}Z_{bd}-\eta_{bd}Z_{ac}+\eta_{ad}Z_{bc},\\
\left[  J_{ab},\tilde{Z}_{cd}\right]   &  =\eta_{bc}\tilde{Z}_{ad}-\eta
_{ac}\tilde{Z}_{bd}-\eta_{bd}\tilde{Z}_{ac}+\eta_{ad}\tilde{Z}_{bc},\\
\left[  \tilde{Z}_{ab},P_{c}\right]   &  =\eta_{bc}\tilde{Z}_{a}-\eta
_{ac}\tilde{Z}_{b},\text{ \ \ \ \ }\left[  \tilde{Z}_{ab},\tilde{Z}%
_{c}\right]  =\eta_{bc}Z_{a}-\eta_{ac}Z_{b}\\
\left[  J_{ab},\tilde{Z}_{c}\right]   &  =\eta_{bc}\tilde{Z}_{a}-\eta
_{ac}\tilde{Z}_{b},
\end{align}%
\begin{align}
\left[  J_{ab},Q_{\alpha}\right]   &  =-\frac{1}{2}\left(  \gamma
_{ab}Q\right)  _{\alpha},\text{ \ \ \ \ \ \ }\left[  J_{ab},\Sigma_{\alpha
}\right]  =-\frac{1}{2}\left(  \gamma_{ab}\Sigma\right)  _{\alpha},\\
\left[  J_{ab},\Phi_{\alpha}\right]   &  =-\frac{1}{2}\left(  \gamma_{ab}%
\Phi\right)  _{\alpha},\text{ \ \ \ \ \ \ }\left[  \tilde{Z}_{ab},Q_{\alpha
}\right]  =-\frac{1}{2}\left(  \gamma_{ab}\Sigma\right)  _{\alpha},\\
\left[  \tilde{Z}_{ab},\Sigma_{\alpha}\right]   &  =-\frac{1}{2}\left(
\gamma_{ab}\Phi\right)  _{\alpha},\text{ \ \ \ \ \ \ }\left[  Z_{ab}%
,Q_{\alpha}\right]  =-\frac{1}{2}\left(  \gamma_{ab}\Phi\right)  _{\alpha},\\
\left[  P_{a},Q_{\alpha}\right]   &  =-\frac{1}{2}\left(  \gamma_{a}%
\Sigma\right)  _{\alpha},\text{ \ \ \ \ \ \ }\left[  P_{a},\Sigma_{\alpha
}\right]  =-\frac{1}{2}\left(  \gamma_{a}\Phi\right)  _{\alpha},\\
\text{ \ \ \ \ \ \ }\left[  \tilde{Z}_{a},Q_{\alpha}\right]   &  =-\frac{1}%
{2}\left(  \gamma_{a}\Phi\right)  _{\alpha},\\
\left\{  Q_{\alpha},Q_{\beta}\right\}   &  =-\frac{1}{2}\left[  \left(
\gamma^{ab}C\right)  _{\alpha\beta}\tilde{Z}_{ab}-2\left(  \gamma^{a}C\right)
_{\alpha\beta}P_{a}\right]  ,\\
\left\{  Q_{\alpha},\Sigma_{\beta}\right\}   &  =-\frac{1}{2}\left[  \left(
\gamma^{ab}C\right)  _{\alpha\beta}Z_{ab}-2\left(  \gamma^{a}C\right)
_{\alpha\beta}\tilde{Z}_{a}\right]  ,\\
\left\{  Q_{\alpha},\Phi_{\beta}\right\}   &  =\left(  \gamma^{a}C\right)
_{\alpha\beta}Z_{a}=\left\{  \Sigma_{\alpha},\Sigma_{\beta}\right\}  ,\\
\text{others}  &  =0,
\end{align}
where we have used the multiplication law of the semigroup $\left(
\ref{lm02}\right)  $ and the commutation relations of the original
superalgebra $\left(  \ref{ads1}\right)  -\left(  \ref{ads5}\right)  $. \ The
new superalgebra obtained after a reduced resonant $S$-expansion of
$\mathfrak{osp}\left(  4|1\right)  $ superalgebra corresponds to a minimal
Maxwell superalgebra type $s\mathcal{M}_{5}$ in $D=4$. \ Interestingly, this
new superalgebra contains the Maxwell algebra type $\mathcal{M}_{5}=\left\{
J_{ab},P_{a},Z_{ab},Z_{a}\right\}  $ as a subalgebra \cite{8, 9}.

In this case, the $S$-expansion method produces two new Majorana spinors
charge $\Sigma$ and $\Phi$. \ These fermionic generators transform as spinors
under Lorentz transformations.\ One sees that the minimal superMaxwell type
$s\mathcal{M}_{5}$ requires new bosonic generators $\left(  \tilde{Z}%
_{ab},\tilde{Z}_{a},Z_{a}\right)  $ and $\Sigma$ is not abelian anymore. \ It
is important to note that setting $\tilde{Z}_{ab}$ and $\tilde{Z}_{a}$ equal
to zero does not lead to a subalgebra. \ In fact, these generators are
required in Jacobi identity for $\left(  Q_{\alpha},Q_{\beta},\Sigma_{\gamma
}\right)  $ due to the gamma matrix identity $\left(  C\gamma^{a}\right)
_{\left(  \alpha\beta\right.  }\left(  C\gamma_{a}\right)  _{\left.
\gamma\delta\right)  }=\left(  C\gamma^{a\beta}\right)  _{\left(  \alpha
\beta\right.  }\left(  C\gamma_{a\beta}\right)  _{\left.  \gamma\delta\right)
}=0$ $\left(  \text{cyclic permutations of }\alpha,\beta,\gamma\right)  $.

It would be interesting to study this algebraic structure in the context of
supergravity theory. \ It seems that the new minimal Maxwell superalgebra
$s\mathcal{M}_{5}$ defined here may enlarge the $D=4$ pure supergravity
lagrangian in a particular way.

\bigskip

\subsection{Minimal $D=4$ superMaxwell algebra type $s\mathcal{M}_{m+2}$}

\bigskip

\qquad Let us further generalize the previous setting. \ In order to obtain
the minimal $D=4$ superMawell algebra type $s\mathcal{M}_{m+2}$, it is
necessary to consider a bigger semigroup. \ Let us consider $S_{E}^{\left(
2m\right)  }=\left\{  \lambda_{0},\lambda_{1},\lambda_{2},\cdots
,\lambda_{2m+1}\right\}  $ as the relevant finite abelian semigroup whose
elements\ are dimensionless and obey the multiplication law%
\begin{equation}
\lambda_{\alpha}\lambda_{\beta}=\left\{
\begin{array}
[c]{c}%
\lambda_{\alpha+\beta}\text{, \ \ \ \ \ \ \ \ when }\alpha+\beta\leq
\lambda_{2m+1},\\
\lambda_{2m+1}\text{, \ \ \ \ \ \ \ when }\alpha+\beta>\lambda_{2m+1}.
\end{array}
\right.  \label{lm03}%
\end{equation}
where $\lambda_{2m+1}$ plays the role of the zero element of the semigroup.
\ \ Let us consider the decomposition $S_{E}^{\left(  2m\right)  }=S_{0}\cup
S_{1}\cup S_{2},$ where the subsets $S_{0},S_{1},S_{2}$ are given by the
general expression%
\begin{equation}
S_{p}=\left\{  \lambda_{2n+p}\text{, with }n=0,\cdots,\left[  \frac{2m-p}%
{2}\right]  \right\}  \cup\left\{  \lambda_{2m+1}\right\}  ,\text{
\ \ \ \ }p=0,1,2.
\end{equation}

This decomposition is said to be resonant since it satisfies [compare with
eqs. $\left(  \ref{ec04}\right)  -\left(  \ref{ec09}\right)  $]%
\begin{align}
S_{0}\cdot S_{0}\subset S_{0},\text{ \ \ \ \ \ \ \ }  &  S_{1}\cdot
S_{1}\subset S_{0}\cap S_{2},\\
S_{0}\cdot S_{1}\subset S_{1},\text{ \ \ \ \ \ \ \ }  &  S_{1}\cdot
S_{2}\subset S_{1},\\
S_{0}\cdot S_{2}\subset S_{2},\text{ \ \ \ \ \ \ \ }  &  S_{2}\cdot
S_{2}\subset S_{0}.
\end{align}

Therefore, we have%
\begin{equation}
\mathfrak{G}_{R}=W_{0}\oplus W_{1}\oplus W_{2},
\end{equation}
with%
\begin{equation}
W_{p}=S_{p}\times V_{p},
\end{equation}
is a resonant subalgebra of $\mathfrak{G}=S\times\mathfrak{g}$.

As in previous cases, it is possible to extract a smaller algebra from the
resonant subalgebra $\mathfrak{G}_{R}$ using the reduction procedure. \ Let
$S_{p}=\hat{S}_{p}\cup\check{S}_{p}$ be a partition of the subsets
$S_{p}\subset S$ where%
\begin{align}
\check{S}_{0}  &  =\left\{  \lambda_{2n}\text{, with }n=0,\cdots,2\left[
m/2\right]  \right\}  ,\text{ \ \ }\hat{S}_{0}=\left\{  \left(  \lambda
_{2m}\right)  ,\lambda_{2m+1}\right\}  ,\text{ \ \ \ \ }\\
\check{S}_{1}  &  =\left\{  \lambda_{2n+1},\text{ with }n=0,\cdots
,m-1\right\}  ,\text{ \ \ }\hat{S}_{1}=\left\{  \lambda_{2m+1}\right\}  ,\\
\check{S}_{2}  &  =\left\{  \lambda_{2n+2},\text{ with }n=0,\cdots,2\left[
\left(  m-1\right)  /2\right]  \right\}  ,\text{ \ \ }\hat{S}_{2}=\left\{
\left(  \lambda_{2m}\right)  ,\lambda_{2m+1}\right\}  ,
\end{align}
where $\left(  \lambda_{2m}\right)  $ means that $\lambda_{2m}\in\hat{S}_{0}$
if $m$ is odd and $\lambda_{2m}\in\hat{S}_{2}$ if $m$ is even. \ For each $p$,
$\hat{S}_{p}\cap\check{S}_{p}=\varnothing$, and using the product $\left(
\ref{lm03}\right)  $ one sees that the partition satisfies [compare with ecs.
$\left(  \ref{ec04}\right)  -\left(  \ref{ec09}\right)  $]%
\begin{equation}%
\begin{tabular}
[c]{ll}%
$\check{S}_{0}\cdot\hat{S}_{0}\subset\hat{S}_{0},$ & $\check{S}_{1}\cdot
\hat{S}_{1}\subset\hat{S}_{0}\cap\hat{S}_{2},$\\
$\check{S}_{0}\cdot\hat{S}_{1}\subset\hat{S}_{1},$ & $\check{S}_{1}\cdot
\hat{S}_{2}\subset\hat{S}_{1},$\\
$\check{S}_{0}\cdot\hat{S}_{2}\subset\hat{S}_{2},$ & $\check{S}_{2}\cdot
\hat{S}_{2}\subset\hat{S}_{0}.$%
\end{tabular}
\end{equation}
Therefore%
\begin{equation}
\mathfrak{\check{G}}_{R}=\check{W}_{0}\oplus\check{W}_{1}\oplus\check{W}_{2},
\end{equation}
corresponds to a reduced algebra of $\mathfrak{G}_{R}$, where%
\begin{align}
\check{W}_{0}  &  =\left(  \check{S}_{0}\times V_{0}\right)  =\left\{
\lambda_{2n}\text{, with }n=0,\cdots,2\left[  m/2\right]  \right\}
\times\left\{  \tilde{J}_{ab}\right\}  ,\\
\check{W}_{1}  &  =\left(  \check{S}_{1}\times V_{1}\right)  =\left\{
\lambda_{2n+1},\text{ with }n=0,\cdots,m-1\right\}  \times\left\{  \tilde
{Q}_{\alpha}\right\}  ,\\
\check{W}_{2}  &  =\left(  \check{S}_{2}\times V_{2}\right)  =\left\{
\lambda_{2n+2},\text{ with }n=0,\cdots,2\left[  \left(  m-1\right)  /2\right]
\right\}  \times\left\{  \tilde{P}_{a}\right\}  .
\end{align}
Here, $\tilde{J}_{ab},\tilde{P}_{a}$ and $\tilde{Q}_{\alpha}$ correspond to
the generators of $\mathfrak{osp}\left(  4|1\right)  $ superalgebra. \ The new
superalgebra obtained by the $S$-expansion procedure is generated by%
\begin{equation}
\left\{  J_{ab},P_{a},Z_{ab}^{\left(  k\right)  },\tilde{Z}_{ab}^{\left(
k\right)  },Z_{a}^{\left(  l\right)  },\tilde{Z}_{a}^{\left(  l\right)
},Q_{\alpha},\Sigma_{\alpha}^{\left(  p\right)  }\right\}  ,
\end{equation}
where these new generators can be written as%
\begin{equation}%
\begin{tabular}
[c]{ll}%
$J_{ab}=J_{ab,0}=\lambda_{0}\tilde{J}_{ab},$ & $\tilde{Z}_{a}^{\left(
l\right)  }=P_{a,4l}=\lambda_{4l}\tilde{P}_{a},$\\
$P_{a}=P_{a,2}=\lambda_{2}\tilde{P}_{a},$ & $Q_{\alpha}=Q_{\alpha,1}%
=\lambda_{1}\tilde{Q}_{\alpha},$\\
$Z_{ab}^{\left(  k\right)  }=J_{ab,4k}=\lambda_{4k}\tilde{J}_{ab},$ &
$\Sigma_{\alpha}^{\left(  k\right)  }=Q_{\alpha,4k-1}=\lambda_{4k-1}\tilde
{Q}_{\alpha},$\\
$\tilde{Z}_{ab}^{\left(  k\right)  }=J_{ab,4k-2}=\lambda_{4k-2}\tilde{J}%
_{ab},$ & $\Phi_{\alpha}^{\left(  l\right)  }=Q_{\alpha,4l+1}=\lambda
_{4l+1}\tilde{Q}_{\alpha},$\\
$Z_{a}^{\left(  l\right)  }=P_{a,4l+2}=\lambda_{4l+2}\tilde{P}_{a}.$ &
\end{tabular}
\end{equation}
with $k=1,\dots,\left[  \frac{m}{2}\right]  $, $l=1,\dots,\left[  \frac
{m-1}{2}\right]  $. \ It is important to note that the super indices $k$ and
$l$ of spinor generators correspond to the expansion labels and they do not
define an $N$-extended superalgebra. The $N$-extended case will be considered
in the next section.

These new generators satisfy the commutation relations%
\begin{align}
\left[  J_{ab},J_{cd}\right]   &  =\eta_{bc}J_{ad}-\eta_{ac}J_{bd}-\eta
_{bd}J_{ac}+\eta_{ad}J_{bc},\label{SMm+2a}\\
\left[  J_{ab},P_{c}\right]   &  =\eta_{bc}P_{a}-\eta_{ac}P_{b},\text{
\ \ \ \ \ \ }\left[  P_{a},P_{b}\right]  =Z_{ab}^{\left(  1\right)  },\\
\left[  J_{ab},Z_{cd}^{\left(  k\right)  }\right]   &  =\eta_{bc}%
Z_{ad}^{\left(  k\right)  }-\eta_{ac}Z_{bd}^{\left(  k\right)  }-\eta
_{bd}Z_{ac}^{\left(  k\right)  }+\eta_{ad}Z_{bc}^{\left(  k\right)  },\\
\left[  Z_{ab}^{\left(  k\right)  },P_{c}\right]   &  =\eta_{bc}Z_{a}^{\left(
k\right)  }-\eta_{ac}Z_{b}^{\left(  k\right)  },\text{ \ \ \ \ \ \ }\left[
J_{ab},Z_{c}^{\left(  l\right)  }\right]  =\eta_{bc}Z_{a}^{\left(  l\right)
}-\eta_{ac}Z_{b}^{\left(  l\right)  },\\
\left[  Z_{ab}^{\left(  k\right)  },Z_{c}^{\left(  l\right)  }\right]   &
=\eta_{bc}Z_{a}^{\left(  k+l\right)  }-\eta_{ac}Z_{b}^{\left(  k+l\right)
},\\
\left[  Z_{ab}^{\left(  k\right)  },Z_{cd}^{\left(  j\right)  }\right]   &
=\eta_{bc}Z_{ad}^{\left(  k+j\right)  }-\eta_{ac}Z_{bd}^{\left(  k+j\right)
}-\eta_{bd}Z_{ac}^{\left(  k+j\right)  }+\eta_{ad}Z_{bc}^{\left(  k+j\right)
},\\
\left[  P_{a},Z_{c}^{\left(  k\right)  }\right]   &  =Z_{ab}^{\left(
k+1\right)  },\text{ \ \ \ \ \ \ }\left[  Z_{a}^{\left(  l\right)  }%
,Z_{c}^{\left(  n\right)  }\right]  =Z_{ab}^{\left(  l+n+1\right)  }
\label{SMm+2b}%
\end{align}%
\begin{align}
\left[  \tilde{Z}_{ab}^{\left(  k\right)  },\tilde{Z}_{cd}^{\left(  j\right)
}\right]   &  =\eta_{bc}Z_{ad}^{\left(  k+j-1\right)  }-\eta_{ac}%
Z_{bd}^{\left(  k+j-1\right)  }-\eta_{bd}Z_{ac}^{\left(  k+j-1\right)  }%
+\eta_{ad}Z_{bc}^{\left(  k+j-1\right)  },\label{SMm+2c}\\
\left[  J_{ab},\tilde{Z}_{cd}^{\left(  k\right)  }\right]   &  =\eta
_{bc}\tilde{Z}_{ad}^{\left(  k\right)  }-\eta_{ac}\tilde{Z}_{bd}^{\left(
k\right)  }-\eta_{bd}\tilde{Z}_{ac}^{\left(  k\right)  }+\eta_{ad}\tilde
{Z}_{bc}^{\left(  k\right)  },\\
\left[  \tilde{Z}_{ab}^{\left(  k\right)  },P_{c}\right]   &  =\eta_{bc}%
\tilde{Z}_{a}^{\left(  k\right)  }-\eta_{ac}\tilde{Z}_{b}^{\left(  k\right)
},\text{ \ \ \ \ \ \ }\left[  J_{ab},\tilde{Z}_{c}^{\left(  l\right)
}\right]  =\eta_{bc}\tilde{Z}_{a}^{\left(  l\right)  }-\eta_{ac}\tilde{Z}%
_{b}^{\left(  l\right)  },\\
\left[  Z_{ab}^{\left(  k\right)  },\tilde{Z}_{c}^{\left(  l\right)  }\right]
&  =\eta_{bc}\tilde{Z}_{a}^{\left(  k+l\right)  }-\eta_{ac}\tilde{Z}%
_{b}^{\left(  k+l\right)  },\text{\ \ \ \ }\left[  \tilde{Z}_{ab}^{\left(
k\right)  },Z_{c}^{\left(  l\right)  }\right]  =\eta_{bc}\tilde{Z}%
_{a}^{\left(  k+l\right)  }-\eta_{ac}\tilde{Z}_{b}^{\left(  k+l\right)  },\\
\left[  \tilde{Z}_{ab}^{\left(  k\right)  },\tilde{Z}_{c}^{\left(  l\right)
}\right]   &  =\eta_{bc}Z_{a}^{\left(  k+l-1\right)  }-\eta_{ac}Z_{b}^{\left(
k+l-1\right)  },\text{ \ \ \ \ \ \ }\left[  P_{a},\tilde{Z}_{b}^{\left(
l\right)  }\right]  =\tilde{Z}_{ab}^{\left(  l+1\right)  }\\
\left[  \tilde{Z}_{a}^{\left(  l\right)  },\tilde{Z}_{b}^{\left(  n\right)
}\right]   &  =Z_{ab}^{\left(  l+n\right)  },\text{ \ \ \ \ \ \ }\left[
Z_{a}^{\left(  l\right)  },\tilde{Z}_{b}^{\left(  n\right)  }\right]
=\tilde{Z}_{ab}^{\left(  l+n+1\right)  },\\
\left[  Z_{ab}^{\left(  k\right)  },\tilde{Z}_{cd}^{\left(  j\right)
}\right]   &  =\eta_{bc}\tilde{Z}_{ad}^{\left(  k+j\right)  }-\eta_{ac}%
\tilde{Z}_{bd}^{\left(  k+j\right)  }-\eta_{bd}\tilde{Z}_{ac}^{\left(
k+j\right)  }+\eta_{ad}\tilde{Z}_{bc}^{\left(  k+j\right)  }, \label{SMm+2d}%
\end{align}%
\begin{align}
\left[  J_{ab},Q_{\alpha}\right]   &  =-\frac{1}{2}\left(  \gamma
_{ab}Q\right)  _{\alpha},\text{ \ \ \ \ \ \ }\left[  J_{ab},\Sigma_{\alpha
}^{\left(  k\right)  }\right]  =-\frac{1}{2}\left(  \gamma_{ab}\Sigma^{\left(
k\right)  }\right)  _{\alpha},\\
\left[  J_{ab},\Phi_{\alpha}^{\left(  l\right)  }\right]   &  =-\frac{1}%
{2}\left(  \gamma_{ab}\Phi^{\left(  l\right)  }\right)  _{\alpha},\text{
\ \ \ \ \ \ }\left[  \tilde{Z}_{ab}^{\left(  k\right)  },Q_{\alpha}\right]
=-\frac{1}{2}\left(  \gamma_{ab}\Sigma^{\left(  k\right)  }\right)  _{\alpha
},\\
\left[  \tilde{Z}_{ab}^{\left(  k\right)  },\Sigma_{\alpha}^{\left(  j\right)
}\right]   &  =-\frac{1}{2}\left(  \gamma_{ab}\Phi^{\left(  k+j-1\right)
}\right)  _{\alpha},\text{ \ \ }\left[  \tilde{Z}_{ab}^{\left(  k\right)
},\Phi_{\alpha}^{\left(  l\right)  }\right]  =-\frac{1}{2}\left(  \gamma
_{ab}\Sigma^{\left(  k+l\right)  }\right)  _{\alpha},\\
\left[  Z_{ab}^{\left(  k\right)  },Q_{\alpha}\right]   &  =-\frac{1}%
{2}\left(  \gamma_{ab}\Phi^{\left(  k\right)  }\right)  _{\alpha},\text{
\ \ \ \ \ \ }\left[  Z_{ab}^{\left(  k\right)  },\Sigma_{\alpha}^{\left(
j\right)  }\right]  =-\frac{1}{2}\left(  \gamma_{ab}\Sigma^{\left(
k+j\right)  }\right)  _{\alpha},\\
\left[  Z_{ab}^{\left(  k\right)  },\Phi_{\alpha}^{\left(  l\right)  }\right]
&  =-\frac{1}{2}\left(  \gamma_{ab}\Phi^{\left(  k+l\right)  }\right)
_{\alpha},\text{ \ \ \ \ \ \ }\left[  P_{a},Q_{\alpha}\right]  =-\frac{1}%
{2}\left(  \gamma_{a}\Sigma^{\left(  1\right)  }\right)  _{\alpha}\\
\left[  P_{a},\Sigma_{\alpha}^{\left(  k\right)  }\right]   &  =-\frac{1}%
{2}\left(  \gamma_{a}\Phi^{\left(  k\right)  }\right)  _{\alpha},\text{
\ \ \ \ \ \ }\left[  P_{a},\Phi_{\alpha}^{\left(  l\right)  }\right]
=-\frac{1}{2}\left(  \gamma_{a}\Sigma^{\left(  l+1\right)  }\right)  _{\alpha
},\\
\left[  \tilde{Z}_{a}^{\left(  l\right)  },Q_{\alpha}\right]   &  =-\frac
{1}{2}\left(  \gamma_{a}\Phi^{\left(  l\right)  }\right)  _{\alpha},\text{
\ \ \ \ \ \ }\left[  \tilde{Z}_{a}^{\left(  l\right)  },\Sigma_{\alpha
}^{\left(  k\right)  }\right]  =-\frac{1}{2}\left(  \gamma_{a}\Sigma^{\left(
l+k\right)  }\right)  _{\alpha},\\
\left[  \tilde{Z}_{a}^{\left(  l\right)  },\Phi_{\alpha}^{\left(  n\right)
}\right]   &  =-\frac{1}{2}\left(  \gamma_{a}\Phi^{\left(  l+n\right)
}\right)  _{\alpha},\text{ \ \ \ \ \ }\left[  Z_{a}^{\left(  l\right)
},Q_{\alpha}\right]  =-\frac{1}{2}\left(  \gamma_{a}\Sigma^{\left(
l+1\right)  }\right)  _{\alpha},\\
\left[  Z_{a}^{\left(  l\right)  },\Sigma_{\alpha}^{\left(  n\right)
}\right]   &  =-\frac{1}{2}\left(  \gamma_{a}\Phi^{\left(  l+n\right)
}\right)  _{\alpha},\text{ \ \ \ \ }\left[  Z_{a}^{\left(  l\right)  }%
,\Phi_{\alpha}^{\left(  n\right)  }\right]  =-\frac{1}{2}\left(  \gamma
_{a}\Sigma^{\left(  l+n+1\right)  }\right)  _{\alpha}, \label{SMm+2e}%
\end{align}%
\begin{align}
\left\{  Q_{\alpha},Q_{\beta}\right\}   &  =-\frac{1}{2}\left[  \left(
\gamma^{ab}C\right)  _{\alpha\beta}\tilde{Z}_{ab}^{\left(  1\right)
}-2\left(  \gamma^{a}C\right)  _{\alpha\beta}P_{a}\right]  ,\label{SMm+2fa}\\
\left\{  Q_{\alpha},\Sigma_{\beta}^{\left(  k\right)  }\right\}   &
=-\frac{1}{2}\left[  \left(  \gamma^{ab}C\right)  _{\alpha\beta}%
Z_{ab}^{\left(  k\right)  }-2\left(  \gamma^{a}C\right)  _{\alpha\beta}%
\tilde{Z}_{a}^{\left(  k\right)  }\right]  ,\\
\left\{  Q_{\alpha},\Phi_{\beta}^{\left(  l\right)  }\right\}   &  =-\frac
{1}{2}\left[  \left(  \gamma^{ab}C\right)  _{\alpha\beta}\tilde{Z}%
_{ab}^{\left(  l+1\right)  }-2\left(  \gamma^{a}C\right)  _{\alpha\beta}%
Z_{a}^{\left(  l\right)  }\right]  ,\\
\left\{  \Sigma_{\alpha}^{\left(  k\right)  },\Sigma_{\beta}^{\left(
j\right)  }\right\}   &  =-\frac{1}{2}\left[  \left(  \gamma^{ab}C\right)
_{\alpha\beta}\tilde{Z}_{ab}^{\left(  k+j\right)  }-2\left(  \gamma
^{a}C\right)  _{\alpha\beta}Z_{a}^{\left(  k+j-1\right)  }\right]  ,\\
\left\{  \Sigma_{\alpha}^{\left(  k\right)  },\Phi_{\beta}^{\left(  l\right)
}\right\}   &  =-\frac{1}{2}\left[  \left(  \gamma^{ab}C\right)  _{\alpha
\beta}Z_{ab}^{\left(  k+l\right)  }-2\left(  \gamma^{a}C\right)  _{\alpha
\beta}\tilde{Z}_{a}^{\left(  k+l\right)  }\right]  ,\\
\left\{  \Phi_{\alpha}^{\left(  l\right)  },\Phi_{\beta}^{\left(  n\right)
}\right\}   &  =-\frac{1}{2}\left[  \left(  \gamma^{ab}C\right)  _{\alpha
\beta}\tilde{Z}_{ab}^{\left(  l+n+1\right)  }-2\left(  \gamma^{a}C\right)
_{\alpha\beta}Z_{a}^{\left(  l+n\right)  }\right]  , \label{SMm+2fb}%
\end{align}
with $k,j=1,\dots,\left[  \frac{m}{2}\right]  $, $l,n=1,\dots,\left[
\frac{m-1}{2}\right]  $. \ The commutation relations can be obtained using the
multiplication law of the semigroup $\left(  \ref{lm03}\right)  $ and the
commutation relations of the original superalgebra $\left(  \ref{ads1}\right)
-\left(  \ref{ads5}\right)  $. \ One sees that when $k+l>\left[  \frac{m}%
{2}\right]  $ then the generatos $T_{A}^{\left(  k\right)  }$ and
$T_{B}^{\left(  l\right)  }$ are abelian.

The new superalgebra obtained after a reduced resonant $S$-expansion of
$\mathfrak{osp}\left(  4|1\right)  $ superalgebra corresponds to the $D=4$
minimal Maxwell superalgebra type $s\mathcal{M}_{m+2}$. \ This superalgebra
contains the Maxwell algebra type $\mathcal{M}_{m+2}=\left\{  J_{ab}%
,P_{a},Z_{ab}^{\left(  k\right)  },Z_{a}^{\left(  l\right)  }\right\}  $ as a
subalgebra $\left(  \text{eqs. }\left(  \ref{SMm+2a}\right)  -\left(
\ref{SMm+2b}\right)  \right)  $ \cite{8, 9}. \ Interestingly, when $m=2$ and
imposing $\tilde{Z}_{ab}^{\left(  1\right)  }=0$ we recover the minimal
Maxwell superalgebra $s\mathcal{M}$. \ The case $m=1$ corresponds to $D=4$
Poincar\'{e} superalgebra $s\mathcal{P}=\left\{  J_{ab},P_{a},Q_{\alpha
}\right\}  $. \ This is not a surprise since the reduced resonant
$S_{E}^{\left(  2\right)  }$-expansion of $\mathfrak{osp}\left(  4|1\right)  $
coincides with a Inon\"{u}-Wigner contraction.

In this case, the $S$-expansion method produces new Majorana spinors charge
$\Sigma^{\left(  k\right)  }$ and $\Phi^{\left(  l\right)  }$. \ These
fermionic generators transform as spinors under Lorentz transformations. \ One
can see that the Jacobi identities for spinors generators are satisfied due to
the gamma matrix identity $\left(  C\gamma^{a}\right)  _{\left(  \alpha
\beta\right.  }\left(  C\gamma_{a}\right)  _{\left.  \gamma\delta\right)
}=\left(  C\gamma^{a\beta}\right)  _{\left(  \alpha\beta\right.  }\left(
C\gamma_{a\beta}\right)  _{\left.  \gamma\delta\right)  }=0$ $\left(
\text{cyclic permutations of }\alpha,\beta,\gamma\right)  $. \ In fact, all
the commutators satisfy the JI since they correspond to expansions of the
original JI of $\mathfrak{osp}\left(  4|1\right)  $.

\bigskip

\section{S-expansion of the $\mathfrak{osp}\left(  4|N\right)  $ superalgebra}

\bigskip

\subsection{$N$-extended superMaxwell algebras}

\bigskip

\qquad We have shown that the minimal $D=4$ Maxwell superalgebras type
$s\mathcal{M}_{m+2}$ can be obtained from a reduced resonant $S_{E}^{\left(
2m\right)  }$-expansion of $\mathfrak{osp}\left(  4|1\right)  $ superalgebra.
\ It seems natural to expect to obtain the $D=4$ $N$-extended Maxwell
superalgebras from an $S$-expansion of $\mathfrak{osp}\left(  4|N\right)  $
superalgebra. \

If we want to apply an $S$-expansion, first it is convenient to decompose the
original superalgebra $\mathfrak{g}$ as a direct sum of subspaces $V_{p}$,%
\begin{align}
\mathfrak{g}=\mathfrak{osp}\left(  4|N\right)   &  =\left(  \mathfrak{so}%
\left(  3,1\right)  \oplus\mathfrak{so}\left(  N\right)  \right)  \oplus
\frac{\mathfrak{osp}\left(  4|N\right)  }{\mathfrak{sp}\left(  4\right)
\oplus\mathfrak{so}\left(  N\right)  }\oplus\frac{\mathfrak{sp}\left(
4\right)  }{\mathfrak{so}\left(  3,1\right)  }\nonumber\\
&  =V_{0}\oplus V_{1}\oplus V_{2},
\end{align}
where $V_{0}$ corresponds to the subspace generated by Lorentz generators
$\tilde{J}_{ab}$ and by $\frac{N\left(  N-1\right)  }{2}$ internal symmetry
generators $T^{ij}$, $V_{1}$ corresponds to the fermionic subspace generated
by$\ N$ Majorana spinor charges $\tilde{Q}_{\alpha}^{i}$ $\left(
i=1,\cdots,N\text{ ; }\alpha=1,\cdots,4\right)  $ and $V_{2}$ corresponds to
the $AdS$ boost generated by $\tilde{P}_{a}$. \ \ The $\mathfrak{osp}\left(
4|N\right)  $ (anti)commutation relations read%
\begin{align}
\left[  \tilde{J}_{ab},\tilde{J}_{cd}\right]   &  =\eta_{bc}\tilde{J}%
_{ad}-\eta_{ac}\tilde{J}_{bd}-\eta_{bd}\tilde{J}_{ac}+\eta_{ad}\tilde{J}%
_{bc},\label{sads01}\\
\left[  T^{ij},T^{kl}\right]   &  =\delta^{jk}T^{il}-\delta^{ik}T^{jl}%
-\delta^{jl}T^{ik}+\delta^{il}T^{jk},\\
\left[  \tilde{J}_{ab},\tilde{P}_{c}\right]   &  =\eta_{bc}\tilde{P}_{a}%
-\eta_{ac}\tilde{P}_{b},\\
\left[  \tilde{P}_{a},\tilde{P}_{b}\right]   &  =\tilde{J}_{ab},\\
\left[  \tilde{J}_{ab},\tilde{Q}_{\alpha}^{i}\right]   &  =-\frac{1}{2}\left(
\gamma_{ab}\tilde{Q}^{i}\right)  _{\alpha},\text{ \ \ \ \ }\left[  \tilde
{P}_{a},\tilde{Q}_{\alpha}^{i}\right]  =-\frac{1}{2}\left(  \gamma_{a}%
\tilde{Q}^{i}\right)  _{\alpha},\\
\left[  T^{ij},\tilde{Q}_{\alpha}^{k}\right]   &  =\left(  \delta^{jk}%
\tilde{Q}_{\alpha}^{i}-\delta^{ik}\tilde{Q}_{\alpha}^{i}\right)  ,\\
\left\{  \tilde{Q}_{\alpha}^{i},\tilde{Q}_{\beta}^{j}\right\}   &  =-\frac
{1}{2}\delta^{ij}\left[  \left(  \gamma^{ab}C\right)  _{\alpha\beta}\tilde
{J}_{ab}-2\left(  \gamma^{a}C\right)  _{\alpha\beta}\tilde{P}_{a}\right]
+C_{\alpha\beta}T^{ij}, \label{sads07}%
\end{align}
where $i,j,k,l=1,\dots,N$.

The subspace structure may be written as%
\begin{align}
\left[  V_{0},V_{0}\right]   &  \subset V_{0},\label{SMNE1}\\
\left[  V_{0},V_{1}\right]   &  \subset V_{1},\\
\left[  V_{0},V_{2}\right]   &  \subset V_{2},\\
\left[  V_{1},V_{1}\right]   &  \subset V_{0}\oplus V_{2},\\
\left[  V_{1},V_{2}\right]   &  \subset V_{1,}\\
\left[  V_{2},V_{2}\right]   &  \subset V_{0}. \label{SMNE6}%
\end{align}

Let us consider $S_{E}^{\left(  4\right)  }=\left\{  \lambda_{0},\lambda
_{1},\lambda_{2},\lambda_{3},\lambda_{4},\lambda_{5}\right\}  $ as the
relevant finite abelian semigroup whose elements\ are dimensionless and obey
the multiplication law%
\begin{equation}
\lambda_{\alpha}\lambda_{\beta}=\left\{
\begin{array}
[c]{c}%
\lambda_{\alpha+\beta}\text{, \ \ \ \ cuando }\alpha+\beta\leq5,\\
\lambda_{5}\text{, \ \ \ \ \ \ \ cuando }\alpha+\beta>5.
\end{array}
\right.  \label{lm05}%
\end{equation}
In this case, $\lambda_{5}$ plays the role of the zero element of the
semigroup $S_{E}^{\left(  4\right)  }$.

Let $S_{E}^{\left(  4\right)  }=S_{0}\cup S_{1}\cup S_{2}$ be a subset
decomposition of $S_{E}^{\left(  4\right)  }$ with%
\begin{align}
S_{0}  &  =\left\{  \lambda_{0},\lambda_{2},\lambda_{4},\lambda_{5}\right\}
,\\
S_{1}  &  =\left\{  \lambda_{1},\lambda_{3},\lambda_{5}\right\}  ,\\
S_{2}  &  =\left\{  \lambda_{2},\lambda_{4},\lambda_{5}\right\}  ,
\end{align}
This subset decomposition satisfies the resonance condition since we have
[compare with eqs. $\left(  \ref{SMNE1}\right)  -\left(  \ref{SMNE6}\right)
$]%
\begin{equation}%
\begin{tabular}
[c]{ll}%
$S_{0}\cdot S_{0}\subset S_{0},$ & $S_{1}\cdot S_{1}\subset S_{0}\cap S_{2}%
,$\\
$S_{0}\cdot S_{1}\subset S_{1},$ & $S_{1}\cdot S_{2}\subset S_{1},$\\
$S_{0}\cdot S_{2}\subset S_{2},$ & $S_{2}\cdot S_{2}\subset S_{0}.$%
\end{tabular}
\end{equation}

Thus, according to Theorem IV.2 of Ref. \cite{5}, we have that%
\begin{equation}
\mathfrak{G}_{R}=W_{0}\oplus W_{1}\oplus W_{2}\text{,}%
\end{equation}
is a resonant subalgebra of $S_{E}^{\left(  4\right)  }\times\mathfrak{g}$,
where%
\begin{align}
W_{0}  &  =\left(  S_{0}\times V_{0}\right)  =\left\{  \lambda_{0},\lambda
_{2},\lambda_{4},\lambda_{5}\right\}  \times\left\{  \tilde{J}_{ab}%
,T^{ij}\right\} \\
&  =\left\{  \lambda_{0}\tilde{J}_{ab},\lambda_{2}\tilde{J}_{ab},\lambda
_{4}\tilde{J}_{ab},\lambda_{5}\tilde{J}_{ab},\lambda_{0}T^{ij},\lambda
_{2}T^{ij},\lambda_{4}T^{ij},\lambda_{5}T^{ij}\right\}  ,\nonumber\\
W_{1}  &  =\left(  S_{1}\times V_{1}\right)  =\left\{  \lambda_{1},\lambda
_{3},\lambda_{5}\right\}  \times\left\{  \tilde{Q}_{\alpha}\right\}  =\left\{
\lambda_{1}\tilde{Q}_{\alpha},\lambda_{3}\tilde{Q}_{\alpha},\lambda_{5}%
\tilde{Q}_{\alpha}\right\}  ,\\
W_{2}  &  =\left(  S_{2}\times V_{2}\right)  =\left\{  \lambda_{2},\lambda
_{4},\lambda_{5}\right\}  \times\left\{  \tilde{P}_{a}\right\}  =\left\{
\lambda_{2}\tilde{P}_{a},\lambda_{4}\tilde{P}_{a},\lambda_{5}\tilde{P}%
_{a}\right\}  .
\end{align}

Imposing $\lambda_{5}T_{A}=0$, the $0_{S}$-reduced resonant superalgebra is
obtained. \ The new superalgebra is generated by $\left\{  J_{ab},P_{a}%
,Z_{ab},\tilde{Z}_{ab},\tilde{Z}_{a},Q_{\alpha}^{i},\Sigma_{\alpha}^{i}%
,T^{ij},Y^{ij},\tilde{Y}^{ij}\right\}  $ where the new generators can be
written as%
\begin{equation}%
\begin{tabular}
[c]{ll}%
$J_{ab}=J_{ab,0}=\lambda_{0}\tilde{J}_{ab},$ & $Q_{\alpha}^{i}=Q_{\alpha
,1}^{i}=\lambda_{1}\tilde{Q}_{\alpha}^{i},$\\
$P_{a}=P_{a,2}=\lambda_{2}\tilde{P}_{a},$ & $\Sigma_{\alpha}^{i}%
=\Sigma_{\alpha,3}^{i}=\lambda_{3}\tilde{Q}_{\alpha}^{i},$\\
$Z_{ab}=J_{ab,4}=\lambda_{4}\tilde{J}_{ab},$ & $T^{ij}=T_{\text{ },0}%
^{ij}=\lambda_{0}T^{ij},$\\
$\tilde{Z}_{ab}=J_{ab,2}=\lambda_{2}\tilde{J}_{ab},$ & $Y^{ij}=T_{\text{ }%
,4}^{\iota j}=\lambda_{4}T^{ij},$\\
$\tilde{Z}_{a}=P_{a,4}=\lambda_{4}\tilde{P}_{a},$ & $\tilde{Y}^{ij}=T_{\text{
},2}^{ij}=\lambda_{2}T^{ij}.$%
\end{tabular}
\end{equation}

Then using the multiplication law of the semigroup $\left(  \ref{lm05}\right)
$ and the commutations relations of the original superalgebra $\left(
\ref{sads01}\right)  -\left(  \ref{sads07}\right)  $ it is possible to write
the resulting superalgebra as%
\begin{align}
\left[  J_{ab},J_{cd}\right]   &  =\eta_{bc}J_{ad}-\eta_{ac}J_{bd}-\eta
_{bd}J_{ac}+\eta_{ad}J_{bc},\label{SMN01}\\
\left[  J_{ab},P_{c}\right]   &  =\eta_{bc}P_{a}-\eta_{ac}P_{b},\text{
\ \ \ \ \ }\left[  P_{a},P_{b}\right]  =Z_{ab},\\
\left[  J_{ab},Z_{cd}\right]   &  =\eta_{bc}Z_{ad}-\eta_{ac}Z_{bd}-\eta
_{bd}Z_{ac}+\eta_{ad}Z_{bc},\\
\left[  J_{ab},\tilde{Z}_{cd}\right]   &  =\eta_{bc}\tilde{Z}_{ad}-\eta
_{ac}\tilde{Z}_{bd}-\eta_{bd}\tilde{Z}_{ac}+\eta_{ad}\tilde{Z}_{bc},\\
\left[  \tilde{Z}_{ab},\tilde{Z}_{cd}\right]   &  =\eta_{bc}Z_{ad}-\eta
_{ac}Z_{bd}-\eta_{bd}Z_{ac}+\eta_{ad}Z_{bc},\\
\left[  J_{ab},\tilde{Z}_{c}\right]   &  =\eta_{bc}\tilde{Z}_{a}-\eta
_{ac}\tilde{Z}_{b},\\
\left[  \tilde{Z}_{ab},P_{c}\right]   &  =\eta_{bc}\tilde{Z}_{a}-\eta
_{ac}\tilde{Z}_{b}, \label{SMN07}%
\end{align}%
\begin{align}
\left[  T^{ij},T^{kl}\right]   &  =\delta^{jk}T^{il}-\delta^{ik}T^{jl}%
-\delta^{jl}T^{ik}+\delta^{il}T^{jk},\\
\left[  T^{ij},Y^{kl}\right]   &  =\delta^{jk}Y^{il}-\delta^{ik}Y^{jl}%
-\delta^{jl}Y^{ik}+\delta^{il}Y^{jk},\\
\left[  T^{ij},\tilde{Y}^{kl}\right]   &  =\delta^{jk}\tilde{Y}^{il}%
-\delta^{ik}\tilde{Y}^{jl}-\delta^{jl}\tilde{Y}^{ik}+\delta^{il}\tilde{Y}%
^{jk},\\
\left[  \tilde{Y}^{ij},\tilde{Y}^{kl}\right]   &  =\delta^{jk}Y^{il}%
-\delta^{ik}Y^{jl}-\delta^{jl}Y^{ik}+\delta^{il}Y^{jk},
\end{align}%
\begin{align}
\left[  J_{ab},Q_{\alpha}^{i}\right]   &  =-\frac{1}{2}\left(  \gamma
_{ab}Q^{i}\right)  _{\alpha},\text{ \ \ \ \ \ \ }\left[  \tilde{Z}%
_{ab},Q_{\alpha}^{i}\right]  =-\frac{1}{2}\left(  \gamma_{ab}\Sigma
^{i}\right)  _{\alpha},\\
\left[  J_{ab},\Sigma_{\alpha}^{i}\right]   &  =-\frac{1}{2}\left(
\gamma_{ab}\Sigma^{i}\right)  _{\alpha},\text{ \ \ \ \ \ \ }\left[
T^{ij},Q_{\alpha}^{i}\right]  =\left(  \delta^{jk}Q_{\alpha}^{i}-\delta
^{ik}Q_{\alpha}^{i}\right)  ,\\
\left[  T^{ij},\Sigma_{\alpha}^{k}\right]   &  =\left(  \delta^{jk}%
\Sigma_{\alpha}^{i}-\delta^{ik}\Sigma_{\alpha}^{i}\right)  ,\\
\left[  \tilde{Y}^{ij},Q_{\alpha}^{k}\right]   &  =\left(  \delta^{jk}%
\Sigma_{\alpha}^{i}-\delta^{ik}\Sigma_{\alpha}^{i}\right)  ,\\
\left[  P_{a},Q_{\alpha}^{i}\right]   &  =-\frac{1}{2}\left(  \gamma_{a}%
\Sigma^{i}\right)  _{\alpha},
\end{align}%
\begin{align}
\left\{  Q_{\alpha}^{i},Q_{\beta}^{j}\right\}   &  =-\frac{1}{2}\delta
^{ij}\left[  \left(  \gamma^{ab}C\right)  _{\alpha\beta}\tilde{Z}%
_{ab}-2\left(  \gamma^{a}C\right)  _{\alpha\beta}P_{a}\right]  +C_{\alpha
\beta}\tilde{Y}^{ij},\label{NMSA1}\\
\left\{  Q_{\alpha}^{i},\Sigma_{\beta}^{j}\right\}   &  =-\frac{1}{2}%
\delta^{ij}\left[  \left(  \gamma^{ab}C\right)  _{\alpha\beta}Z_{ab}-2\left(
\gamma^{a}C\right)  _{\alpha\beta}\tilde{Z}_{a}\right]  +C_{\alpha\beta}%
Y^{ij},\label{NMSA2}\\
\text{others }  &  =0.
\end{align}
\qquad

The new superalgebra obtained after a reduced resonant $S_{E}^{\left(
4\right)  }$-expansion of $\mathfrak{osp}\left(  4|N\right)  $ superalgebra
corresponds to the $D=4$ $N$-extended Maxwell superalgebra $s\mathcal{M}%
_{4}^{\left(  N\right)  }$. \ An alternative expansion procedure to obtain the
$N$-extended Maxwell superalgebra has been proposed in Ref. \cite{19}.
\ Interestingly, this superalgebra contains the generalized Maxwell algebra
$g\mathcal{M}=\left\{  J_{ab},P_{a},Z_{ab},\tilde{Z}_{ab},\tilde{Z}%
_{a}\right\}  $ as a subalgebra (see Appendix B). \ One sees that the
$S$-expansion procedure introduces additional bosonic generators which modify
the minimal Maxwell superalgebra [see eqs. $\left(  \ref{NMSA1}\right)  $,
$\left(  \ref{NMSA2}\right)  $]. \ Naturally when $\tilde{Z}_{a}=\tilde
{Z}_{ab}=Y^{ij}=\tilde{Y}^{ij}=0$, we obtain the simplest $D=4$ $N$-extended
Maxwell superalgebra $s\mathcal{M}^{\left(  N\right)  }$ generated by
$\left\{  J_{ab},P_{a},Z_{ab},Q_{\alpha}^{i},\Sigma_{\alpha}^{i}%
,T_{ab}\right\}  $ $.$ Eventually for $N=1$, with $T_{ab}=0$, the $D=4$
minimal Maxwell superalgebra $s\mathcal{M}$ is recovered. \ It is important to
note that setting some generators equals to zero does not always lead to a Lie
superalgebra. \ Nevertheless, the properties of the gamma matrices in $4$
dimensions permit us to impose some generators equals to zero without breaking
the Jacobi Identity.

We can generalize this procedure and obtain the $N$-extended superMaxwell
algebra type $s\mathcal{M}_{m+2}^{\left(  N\right)  }$ as an reduced resonant
$S$-expansion of $\mathfrak{osp}\left(  4|N\right)  $ with $S_{E}^{\left(
2m\right)  }=\left\{  \lambda_{0},\lambda_{1},\lambda_{2},\cdots
,\lambda_{2m+1}\right\}  $ as abelian semigroup. \ In fact, if we consider a
resonant subset decomposition $S_{E}^{\left(  2m\right)  }=S_{0}\cup S_{1}\cup
S_{2},$ where%
\begin{equation}
S_{p}=\left\{  \lambda_{2n+p}\text{, with }n=0,\cdots,\left[  \frac{2m-p}%
{2}\right]  \right\}  \cup\left\{  \lambda_{2m+1}\right\}  ,\text{
\ \ \ \ }p=0,1,2,
\end{equation}
and let $S_{p}=\hat{S}_{p}\cup\check{S}_{p}$ be a partition of the subsets
$S_{p}\subset S$ where%
\begin{align}
\check{S}_{0}  &  =\left\{  \lambda_{2n}\text{, with }n=0,\cdots,2\left[
m/2\right]  \right\}  ,\text{ \ \ }\hat{S}_{0}=\left\{  \left(  \lambda
_{2m}\right)  ,\lambda_{2m+1}\right\}  ,\text{ \ \ \ \ }\\
\check{S}_{1}  &  =\left\{  \lambda_{2n+1},\text{ with }n=0,\cdots
,m-1\right\}  ,\text{ \ \ }\hat{S}_{1}=\left\{  \lambda_{2m+1}\right\}  ,\\
\check{S}_{2}  &  =\left\{  \lambda_{2n+2},\text{ with }n=0,\cdots,2\left[
\left(  m-1\right)  /2\right]  \right\}  ,\text{ \ \ }\hat{S}_{2}=\left\{
\left(  \lambda_{2m}\right)  ,\lambda_{2m+1}\right\}  ,
\end{align}
where $\left(  \lambda_{2m}\right)  $ means that $\lambda_{2m}\in\hat{S}_{0}$
if $m$ is odd and $\lambda_{2m}\in\hat{S}_{2}$ if $m$ is even. \ This
decomposition satifies the resonant condition for any value of $m$ and\ we
find that
\begin{equation}
\mathfrak{\check{G}}_{R}=\left(  \check{S}_{0}\times V_{0}\right)
\oplus\left(  \check{S}_{1}\times V_{1}\right)  \oplus\left(  \check{S}%
_{2}\times V_{2}\right)  ,
\end{equation}
corresponds to a reduced resonant algebra. \ This new superalgebra correspond
to the $N$-extended Maxwell superalgebra type $s\mathcal{M}_{m+2}^{\left(
N\right)  }$ which is generated by
\begin{equation}
\left\{  J_{ab},P_{a},Z_{ab}^{\left(  k\right)  },\tilde{Z}_{ab}^{\left(
k\right)  },Z_{a}^{\left(  k\right)  },\tilde{Z}_{a}^{\left(  k\right)
},Q_{\alpha}^{i},\Sigma_{\alpha}^{i\left(  k\right)  },\Phi_{\alpha}^{i\left(
k\right)  },T^{ij},Y^{ij\left(  k\right)  },\tilde{Y}^{ij\left(  k\right)
}\right\}  .
\end{equation}
These generators can be written as%
\begin{equation}%
\begin{tabular}
[c]{ll}%
$J_{ab}=J_{ab,0}=\lambda_{0}\tilde{J}_{ab},$ & $P_{a}=P_{a,2}=\lambda
_{2}\tilde{P}_{a},$\\
$Z_{ab}^{\left(  k\right)  }=J_{ab,4k}=\lambda_{4k}\tilde{J}_{ab},$ &
$\tilde{Z}_{ab}^{\left(  k\right)  }=J_{ab,4k-2}=\lambda_{4k-2}\tilde{J}%
_{ab},$\\
$Z_{a}^{\left(  l\right)  }=P_{a,4l+2}=\lambda_{4l+2}\tilde{P}_{a},\text{ }$ &
$\tilde{Z}_{a}^{\left(  l\right)  }=P_{a,4l}=\lambda_{4l}\tilde{P}_{a},$\\
$Q_{\alpha}^{i}=Q_{\alpha,1}^{i}=\lambda_{1}\tilde{Q}_{\alpha}^{i},$ &
$\Sigma_{\alpha}^{i\left(  k\right)  }=Q_{\alpha,4k-1}^{i}=\lambda
_{4k-1}\tilde{Q}_{\alpha}^{i},$\\
$\Phi_{\alpha}^{i\left(  l\right)  }=Q_{\alpha,4l+1}^{i}=\lambda_{4l+1}%
\tilde{Q}_{\alpha}^{i},$ & $T^{ij}=T_{\text{ },0}^{ij}=\lambda_{0}T^{ij},$\\
$Y^{ij\left(  k\right)  }=T_{\text{ },4k}^{\iota j}=\lambda_{4k}T^{ij},$ &
$\tilde{Y}^{ij\left(  k\right)  }=T_{\text{ },4k-2}^{ij}=\lambda_{4k-2}%
T^{ij},$%
\end{tabular}
\end{equation}
with $k=1,\dots,\left[  \frac{m}{2}\right]  $, $l=1,\dots,\left[  \frac
{m-1}{2}\right]  $, $i,j=1,\dots,N$. \ The new bosonics generators $\left\{
Z_{ab},\tilde{Z}_{ab},Z_{a},\tilde{Z}_{a},Y^{ij},\tilde{Y}^{ij}\right\}  $
modify some anticommutators of the minimal Maxwell superalgebra type ($\left(
\ref{SMm+2fa}\right)  -\left(  \ref{SMm+2fb}\right)  $) . \ Now we have%
\begin{align}
\left\{  Q_{\alpha}^{i},Q_{\beta}^{j}\right\}   &  =-\frac{1}{2}\delta
^{ij}\left[  \left(  \gamma^{ab}C\right)  _{\alpha\beta}\tilde{Z}%
_{ab}^{\left(  1\right)  }-2\left(  \gamma^{a}C\right)  _{\alpha\beta}%
P_{a}\right]  +C_{\alpha\beta}\tilde{Y}^{ij\left(  1\right)  },\\
\left\{  Q_{\alpha}^{i},\Sigma_{\beta}^{j\left(  k\right)  }\right\}   &
=-\frac{1}{2}\delta^{ij}\left[  \left(  \gamma^{ab}C\right)  _{\alpha\beta
}Z_{ab}^{\left(  k\right)  }-2\left(  \gamma^{a}C\right)  _{\alpha\beta}%
\tilde{Z}_{a}^{\left(  k\right)  }\right]  +C_{\alpha\beta}Y^{ij\left(
k\right)  },\\
\left\{  Q_{\alpha}^{i},\Phi_{\beta}^{j\left(  l\right)  }\right\}   &
=-\frac{1}{2}\delta^{ij}\left[  \left(  \gamma^{ab}C\right)  _{\alpha\beta
}\tilde{Z}_{ab}^{\left(  l+1\right)  }-2\left(  \gamma^{a}C\right)
_{\alpha\beta}Z_{a}^{\left(  l\right)  }\right]  +C_{\alpha\beta}\tilde
{Y}^{ij\left(  l+1\right)  },\\
\left\{  \Sigma_{\alpha}^{i\left(  k\right)  },\Sigma_{\beta}^{j\left(
q\right)  }\right\}   &  =-\frac{1}{2}\delta^{ij}\left[  \left(  \gamma
^{ab}C\right)  _{\alpha\beta}\tilde{Z}_{ab}^{\left(  k+q\right)  }-2\left(
\gamma^{a}C\right)  _{\alpha\beta}Z_{a}^{\left(  k+q-1\right)  }\right]
+C_{\alpha\beta}\tilde{Y}^{ij\left(  k+q\right)  },\\
\left\{  \Sigma_{\alpha}^{i\left(  k\right)  },\Phi_{\beta}^{j\left(
l\right)  }\right\}   &  =-\frac{1}{2}\delta^{ij}\left[  \left(  \gamma
^{ab}C\right)  _{\alpha\beta}Z_{ab}^{\left(  k+l\right)  }-2\left(  \gamma
^{a}C\right)  _{\alpha\beta}\tilde{Z}_{a}^{\left(  k+l\right)  }\right]
+C_{\alpha\beta}Y^{ij\left(  k+l\right)  },\\
\left\{  \Phi_{\alpha}^{\left(  l\right)  },\Phi_{\beta}^{\left(  n\right)
}\right\}   &  =-\frac{1}{2}\delta^{ij}\left[  \left(  \gamma^{ab}C\right)
_{\alpha\beta}\tilde{Z}_{ab}^{\left(  l+n+1\right)  }-2\left(  \gamma
^{a}C\right)  _{\alpha\beta}Z_{a}^{\left(  l+n\right)  }\right]
+C_{\alpha\beta}\tilde{Y}^{ij\left(  l+n+1\right)  },
\end{align}
with $k,q=1,\dots,\left[  \frac{m}{2}\right]  $, $l,n=1,\dots,\left[
\frac{m-1}{2}\right]  $, $i,j=1,\dots,N$. \ The internal symmetries generators
also brings some new commutation relations besides the commutators $\left(
\ref{SMm+2a}\right)  -\left(  \ref{SMm+2e}\right)  $,%
\begin{align}
\left[  T^{ij},T^{gh}\right]   &  =\delta^{jg}T^{ih}-\delta^{ig}T^{jh}%
-\delta^{jh}T^{ig}+\delta^{ih}T^{jg},\\
\left[  T^{ij},Y^{gh\left(  k\right)  }\right]   &  =\delta^{jg}Y^{ih\left(
k\right)  }-\delta^{ig}Y^{jh\left(  k\right)  }-\delta^{jh}Y^{ig\left(
k\right)  }+\delta^{ih}Y^{jg\left(  k\right)  },\\
\left[  T^{ij},\tilde{Y}^{gh\left(  k\right)  }\right]   &  =\delta^{jg}%
\tilde{Y}^{ih\left(  k\right)  }-\delta^{ig}\tilde{Y}^{jh\left(  k\right)
}-\delta^{jh}\tilde{Y}^{ig\left(  k\right)  }+\delta^{ih}\tilde{Y}^{jg\left(
k\right)  },\\
\left[  \tilde{Y}^{ij\left(  k\right)  },\tilde{Y}^{gh\left(  q\right)
}\right]   &  =\delta^{jg}Y^{ih\left(  k+q-1\right)  }-\delta^{ig}Y^{jh\left(
k+q-1\right)  }-\delta^{jh}Y^{ig\left(  k+q-1\right)  }+\delta^{ih}%
Y^{jg\left(  k+q-1\right)  },\\
\left[  \tilde{Y}^{ij\left(  k\right)  },Y^{gh\left(  q\right)  }\right]   &
=\delta^{jg}\tilde{Y}^{ih\left(  k+q\right)  }-\delta^{ig}\tilde{Y}^{jh\left(
k+q\right)  }-\delta^{jh}\tilde{Y}^{ig\left(  k+q\right)  }+\delta^{ih}%
\tilde{Y}^{jg\left(  k+q\right)  },\\
\left[  Y^{ij\left(  k\right)  },Y^{gh\left(  q\right)  }\right]   &
=\delta^{jg}Y^{ih\left(  k+q\right)  }-\delta^{ig}Y^{jh\left(  k+q\right)
}-\delta^{jh}Y^{ig\left(  k+q\right)  }+\delta^{ih}Y^{jg\left(  k+q\right)
},\\
\left[  T^{ij},Q_{\alpha}^{i}\right]   &  =\left(  \delta^{jk}Q_{\alpha}%
^{i}-\delta^{ik}Q_{\alpha}^{i}\right)  ,\\
\left[  T^{ij},\Sigma_{\alpha}^{g\left(  k\right)  }\right]   &  =\left[
\tilde{Y}^{ij\left(  k\right)  },Q_{\alpha}^{g}\right]  =\left(  \delta
^{jg}\Sigma_{\alpha}^{i\left(  k\right)  }-\delta^{ig}\Sigma_{\alpha
}^{i\left(  k\right)  }\right)  ,\\
\left[  T^{ij},\Phi_{\alpha}^{g\left(  k\right)  }\right]   &  =\text{\ }%
\left[  Y^{ij\left(  k\right)  },Q_{\alpha}^{g}\right]  =\left(  \delta
^{jg}\Phi_{\alpha}^{i\left(  k\right)  }-\delta^{ig}\Phi_{\alpha}^{i\left(
k\right)  }\right)  ,\\
\left[  \tilde{Y}^{ij\left(  k\right)  },\Phi_{\alpha}^{g\left(  q\right)
}\right]   &  =\left[  Y^{ij\left(  k\right)  },\Sigma_{\alpha}^{g\left(
q\right)  }\right]  =\left(  \delta^{jg}\Sigma_{\alpha}^{i\left(  k+q\right)
}-\delta^{ig}\Sigma_{\alpha}^{i\left(  k+q\right)  }\right)  ,\\
\text{\ }\left[  \tilde{Y}^{ij\left(  k\right)  },\Sigma_{\alpha}^{g\left(
q\right)  }\right]   &  =\left(  \delta^{jg}\Phi_{\alpha}^{i\left(
k+q-1\right)  }-\delta^{ig}\Phi_{\alpha}^{i\left(  k+q-1\right)  }\right)  ,\\
\left[  Y^{ij\left(  k\right)  },\Phi_{\alpha}^{g\left(  q\right)  }\right]
&  =\left(  \delta^{jg}\Phi_{\alpha}^{i\left(  k+q\right)  }-\delta^{ig}%
\Phi_{\alpha}^{i\left(  k+q\right)  }\right)  .
\end{align}

The commutation relations can be obtained using the multiplication law of the
semigroup and the commutation relations of the $\mathfrak{osp}\left(
4|N\right)  $ superalgebra. \ As in the case of minimal superMaxwell algebra
type one sees that when $k+q>\left[  \frac{m}{2}\right]  $ then the generators
$T_{A}^{\left(  k\right)  }$ and $T_{B}^{\left(  q\right)  }$ are abelian.
\ As in the previous case, the $S$-expansion method produces new Majorana
spinors charge $\Sigma^{i\left(  k\right)  }$ and $\Phi^{i\left(  l\right)  }$
which transform as spinors under Lorentz transformations.

The $N$-extended Maxwell superalgebra type $s\mathcal{M}_{m+2}^{\left(
N\right)  }$ contains the Maxwell algebra type $\mathcal{M}_{m+2}=\left\{
J_{ab},P_{a},Z_{ab}^{\left(  k\right)  },Z_{a}^{\left(  l\right)  }\right\}  $
as a subalgebra $\left(  \text{eqs. }\left(  \ref{SMm+2a}\right)  -\left(
\ref{SMm+2b}\right)  \right)  $ \cite{9}. We can see that for $m=2$ we recover
the $D=4$ $N$-extended Maxwell superalgebra $s\mathcal{M}_{4}^{\left(
N\right)  }$. \ It is interesting to observe that for $m=1$ we obtain the
$D=4$ $N$-extended Poincar\'{e} superalgebra $s\mathcal{P}^{\left(  N\right)
}=\left\{  J_{ab},P_{a},Q_{\alpha},T^{ij}\right\}  $. \ \ This is not a
surprise because the reduced resonant $S_{E}^{\left(  2\right)  }$-expansion
of $\mathfrak{osp}\left(  4|N\right)  $ coincides with an Inon\"{u}-Wigner contraction.

Interestingly, it is possible to write the $N$-extended Maxwell superalgebra
type $s\mathcal{M}_{m+2}^{\left(  N\right)  }$ in a very compact way defining%
\begin{align*}
J_{ab,\left(  k\right)  }  &  =\lambda_{2k}\tilde{J}_{ab},\\
P_{a,\left(  l\right)  }  &  =\lambda_{2l}\tilde{P}_{a},\\
Q_{\alpha,\left(  p\right)  }  &  =\lambda_{2p-1}\tilde{Q}_{\alpha},\\
Y_{\left(  k\right)  }^{ij}  &  =\lambda_{2k}T^{ij}%
\end{align*}
with $k=0,\dots,m-1$; $l=1,\dots,m$; $p=1,\dots,m$ when $m$ is odd and
$k=0,\dots,m$; $l=1,\dots,m-1$; $p=1,\dots,m$ when $m$ is even. Here, the
generators $\tilde{J}_{ab},\tilde{P}_{a},\tilde{Q}_{\alpha}$ and $T^{ij}$
correspond to the $\mathfrak{osp}\left(  4|N\right)  $ generators. \ Then
using the multiplication law of the semigroup $\left(  \ref{lm03}\right)  $
and the commutations relations of the original superalgebra $\left(
\ref{sads01}\right)  -\left(  \ref{sads07}\right)  $ is possible to write the
resulting superalgebra as%
\begin{align}
\left[  J_{ab,\left(  k\right)  },J_{cd,\left(  j\right)  }\right]   &
=\eta_{bc}J_{ad,\left(  k+j\right)  }-\eta_{ac}J_{bd,\left(  k+j\right)
}-\eta_{bd}J_{ac,\left(  k+j\right)  }+\eta_{ad}J_{bc,\left(  k+j\right)  },\\
\left[  Y_{\left(  k\right)  }^{ij},Y_{\left(  j\right)  }^{gh}\right]   &
=\delta^{jg}Y_{\left(  k+j\right)  }^{ih}-\delta^{ig}Y_{\left(  k+j\right)
}^{jh}-\delta^{jh}Y_{\left(  k+j\right)  }^{ig}+\delta^{ih}Y_{\left(
k+j\right)  }^{jg},\\
\left[  J_{ab,\left(  k\right)  },P_{c,\left(  l\right)  }\right]   &
=\eta_{bc}P_{a,\left(  k+l\right)  }-\eta_{ac}P_{b,\left(  k+l\right)  },\\
\left[  P_{a,\left(  l\right)  },P_{b,\left(  n\right)  }\right]   &
=J_{ab,\left(  l+n\right)  },\\
\left[  J_{ab,\left(  k\right)  },Q_{\alpha,\left(  p\right)  }\right]   &
=-\frac{1}{2}\left(  \gamma_{ab}Q\right)  _{\alpha,\left(  k+p\right)  },\\
\left[  P_{a,\left(  l\right)  },Q_{\alpha,\left(  p\right)  }\right]   &
=-\frac{1}{2}\left(  \gamma_{a}Q\right)  _{\alpha,\left(  l+p\right)  },\\
\left[  T_{\left(  k\right)  }^{ij},Q_{\alpha,\left(  p\right)  }^{g}\right]
&  =\left(  \delta^{jg}Q_{\alpha,\left(  k+p\right)  }^{i}-\delta
^{ig}Q_{\alpha,\left(  k+p\right)  }^{i}\right)  ,\\
\left\{  Q_{\alpha,\left(  p\right)  },Q_{\beta,\left(  q\right)  }\right\}
&  =-\frac{1}{2}\left[  \left(  \gamma^{ab}C\right)  _{\alpha\beta
}J_{ab,\left(  p+q\right)  }-2\left(  \gamma^{a}C\right)  _{\alpha\beta
}P_{a,\left(  p+q\right)  }\right]  +C_{\alpha\beta}Y_{\left(  p+q\right)
}^{ij},
\end{align}
where $i,j,g,h=1,\cdots,N$. \ Naturally, when $k+j>m$ then the generators
$T_{A}^{\left(  k\right)  }$ and $T_{B}^{\left(  j\right)  }$ are abelian.
With this notation it is not trivial to see the Maxwell algebra type
$\mathcal{M}_{m+2}$ as a subalgebra. \ However it could be useful in order to
construct an action for this superalgebra.

\bigskip

\section{Comments and possible developments}

\bigskip

\qquad In the present work we have shown that\ the Maxwell superalgebras found
by the MC expansion method in Ref. \cite{19} can be derived alternatively by
the $S$-expansion procedure. \ In particular, the $S$-expansion of
$\mathfrak{osp}\left(  4|1\right)  $ permits us to obtain the minimal Maxwell
superalgebra $s\mathcal{M}$. \ Then choosing different semigroups we have
shown that it is possible to define new minimal $D=4$ Maxwell superalgebras
type $s\mathcal{M}_{m+2}$ which can be seen as a generalization of the
D'Auria-Fr\'{e} superalgebra and the Green algebras introduced in Refs.
\cite{7}, \cite{22}, respectively. \ Interestingly, the case $m=1$ corresponds
to the minimal Poincar\'{e} superalgebra. \ Recently it was shown that the
minimal Maxwell superalgebra $s\mathcal{M}$ may be used to obtain the minimal
$D=4$ pure supergravity \cite{21}. \ It seems that the new minimal Maxwell
superalgebras $s\mathcal{M}_{m+2}$ defined here may be good candidates to
enlarge the $D=4$ pure supergravity lagrangian leading to a generalized
cosmological term. \ Interestingly we have shown that this Maxwell
superalgebra contains the Maxwell algebras type $\mathcal{M}_{m+2}$ as bosonic subalgebras.

We also have shown that the $D=4$ $N$-extended Maxwell superalgebra
$s\mathcal{M}^{\left(  N\right)  }$, derived initially as a MC expansion in
Ref. \cite{19}, can be obtained alternatively as an $S$-expansion of
$\mathfrak{osp}\left(  4|N\right)  $. \ In this case the $S$-expansion
produces additional bosonic generators which modify the minimal Maxwell
superalgebra. \ Choosing bigger semigroups we have shown that it is possible
to define new $D=4$ $N$-extended Maxwell superalgebras type $s\mathcal{M}%
_{m+2}^{\left(  N\right)  }$. \ Naturally when $m=2$ we recover the
$s\mathcal{M}^{\left(  N\right)  }$ superalgebra and for $N=1$ we recover the
Maxwell algebra type $s\mathcal{M}_{m+2}$. \ It would be interesting to build
lagrangians with the $2$-form curvature associated to these new $N$-extended
Maxwell superalgebras $s\mathcal{M}_{m+2}^{\left(  N\right)  }$ and study
their relation with the $N$-extended supergravity in $D=4$ [work in progress].

Thus, we have shown that the $S$-expansion procedure is a powerful and simple
tool in order to derive new Lie superalgebras. \ In fact, the introduction of
new Majorana spinor charges could not be guessed trivially. \ The method
considered here could play an important role in the context of supergravity in
higher dimensions. \ It seems that it should be possible to recover standard
odd- and even-dimensional supergravity from the Maxwell superalgebra family
[work in progress].

\section*{Acknowledgements \textbf{\ }}

\qquad The authors wish to thank L. Andrianopoli, R. D' Auria and M. Trigiante
for enlightening discussions and their\ kind hospitality at Dipartimento
Scienza Applicata e Tecnologia of Politecnico di Torino, where this work was
done. \ We are also grateful to P. Salgado for introducing the topics covered
in the present work. \ The autors also thank O. Fierro for helpful comments on
the subject of this work. \ The autors were supported by grants from the
Comisi\'{o}n Nacional de Investigaci\'{o}n Cient\'{\i}fica y Tecnol\'{o}gica
(CONICYT) and from the Universidad de Concepci\'{o}n, Chile. \ This work was
supported in part by FONDECYT Grants N$%
{{}^\circ}%
$ 1130653.

\bigskip

\appendix

\section{\label{appex}$S$\textbf{-expansion of the commutation relations}}

\bigskip

\qquad Let $\mathfrak{g}$ be a Lie (super)algebra given by%
\[
\left[  T_{A},T_{B}\right]  =C_{AB}^{\text{ \ \ }C}T_{C}\text{.}%
\]
Let $S=\left\{  \lambda_{\alpha}\right\}  $ be an abelian semigroup with
$2$-selector $K_{\alpha\beta}^{\text{ \ }\gamma}$. \ Let us denote a basis
element of the direct product $S\times g$ by $T_{\left(  A,\alpha\right)
}=\lambda_{\alpha}T_{A}$ and consider the induced commutator $\left[
T_{\left(  A,\alpha\right)  },T_{\left(  B,\beta\right)  }\right]
=\lambda_{\alpha}\lambda_{\beta}\left[  T_{A},T_{B}\right]  $. \ Then
$S\times\mathfrak{g}$ is also a Lie (super)algebra with structure constants%
\begin{equation}
C_{\left(  A,\alpha\right)  \left(  B,\beta\right)  }^{\text{
\ \ \ \ \ \ \ \ }\left(  C,\gamma\right)  }=K_{\alpha\beta}^{\text{ \ }\gamma
}C_{AB}^{\text{ \ \ }C}\text{.}%
\end{equation}
Let us consider as example the $S_{E}^{\left(  4\right)  }$-expansion of the
anticommutator of $\mathfrak{osp}\left(  4|1\right)  $,%
\begin{equation}
\left\{  \tilde{Q}_{\alpha},\tilde{Q}_{\beta}\right\}  =-\frac{1}{2}\left[
\left(  \gamma^{ab}C\right)  _{\alpha\beta}\tilde{J}_{ab}-2\left(  \gamma
^{a}C\right)  _{\alpha\beta}\tilde{P}_{a}\right]  .
\end{equation}
\ We have said that a decomposition of the original algebra $\mathfrak{g}$ is
given by,%
\begin{align}
\mathfrak{g}=\mathfrak{osp}\left(  4|1\right)   &  =\mathfrak{so}\left(
3,1\right)  \oplus\frac{\mathfrak{osp}\left(  4|1\right)  }{\mathfrak{sp}%
\left(  4\right)  }\oplus\frac{\mathfrak{sp}\left(  4\right)  }{\mathfrak{so}%
\left(  3,1\right)  }\nonumber\\
&  =V_{0}\oplus V_{1}\oplus V_{2},
\end{align}
Let $S_{p}=\hat{S}_{p}\cup\check{S}_{p}$ be a partition of the subsets
$S_{p}\subset S_{E}^{\left(  4\right)  }=\left\{  \lambda_{0},\lambda
_{1},\lambda_{2},\lambda_{3},\lambda_{4},\lambda_{5}\right\}  $ where%
\begin{align}
\check{S}_{0}  &  =\left\{  \lambda_{0},\lambda_{4},\lambda_{2}\right\}
,\text{ \ \ \ \ }\hat{S}_{0}=\left\{  \lambda_{5}\right\}  ,\\
\check{S}_{1}  &  =\left\{  \lambda_{1},\lambda_{3}\right\}  ,\text{
\ \ \ \ }\hat{S}_{1}=\left\{  \lambda_{5}\right\}  ,\\
\check{S}_{2}  &  =\left\{  \lambda_{2}\right\}  ,\text{ \ \ \ \ \ \ \ \ }%
\hat{S}_{2}=\left\{  \lambda_{4},\lambda_{5}\right\}  .
\end{align}
Then, we have said that%
\begin{equation}
\mathfrak{\check{G}}_{R}=\left(  \check{S}_{0}\times V_{0}\right)
\oplus\left(  \check{S}_{1}\times V_{1}\right)  \oplus\left(  \check{S}%
_{2}\times V_{2}\right)  ,
\end{equation}
corresponds to a reduced resonant superalgebra. \ Thus the new Majorana spinor
charges are given by%
\begin{align}
Q_{\alpha}  &  =Q_{\alpha,1}=\lambda_{1}\tilde{Q}_{\alpha},\\
\Sigma_{\alpha}  &  =Q_{\alpha,3}=\lambda_{3}\tilde{Q}_{\alpha},
\end{align}
where $\tilde{Q}_{\alpha}$ corresponds to the original Majorana spinor charge.
\ Then, the new anticommutators are given by%
\begin{align}
\left\{  Q_{\alpha},Q_{\beta}\right\}   &  =\left\{  \lambda_{1}\tilde
{Q}_{\alpha},\lambda_{1}\tilde{Q}_{\beta}\right\} \nonumber\\
&  =\lambda_{1}\lambda_{1}\left\{  \tilde{Q}_{\alpha},\tilde{Q}_{\beta
}\right\} \nonumber\\
&  =-\lambda_{2}\frac{1}{2}\left[  \left(  \gamma^{ab}C\right)  _{\alpha\beta
}\tilde{J}_{ab}-2\left(  \gamma^{a}C\right)  _{\alpha\beta}\tilde{P}%
_{a}\right] \nonumber\\
&  =-\frac{1}{2}\left[  \left(  \gamma^{ab}C\right)  _{\alpha\beta}\lambda
_{2}\tilde{J}_{ab}-2\left(  \gamma^{a}C\right)  _{\alpha\beta}\lambda
_{2}\tilde{P}_{a}\right] \nonumber\\
&  =-\frac{1}{2}\left[  \left(  \gamma^{ab}C\right)  _{\alpha\beta}\tilde
{Z}_{ab}-2\left(  \gamma^{a}C\right)  _{\alpha\beta}P_{a}\right]  ,
\end{align}
where we have used that $\tilde{Z}_{ab}=J_{ab,2}=\lambda_{2}\tilde{J}_{ab}$
and $P_{a}=P_{a,2}=\lambda_{2}\tilde{P}_{a}$. \ In the same way, it is
possible to show that%
\begin{align}
\left\{  Q_{\alpha},\Sigma_{\beta}\right\}   &  =\left\{  \lambda_{1}\tilde
{Q}_{\alpha},\lambda_{3}\tilde{Q}_{\beta}\right\} \nonumber\\
&  =-\lambda_{4}\frac{1}{2}\left[  \left(  \gamma^{ab}C\right)  _{\alpha\beta
}\tilde{J}_{ab}-2\left(  \gamma^{a}C\right)  _{\alpha\beta}\tilde{P}%
_{a}\right] \nonumber\\
&  =-\frac{1}{2}\left[  \left(  \gamma^{ab}C\right)  _{\alpha\beta}\lambda
_{4}\tilde{J}_{ab}-2\left(  \gamma^{a}C\right)  _{\alpha\beta}\lambda
_{4}\tilde{P}_{a}\right] \nonumber\\
&  =-\frac{1}{2}\left[  \left(  \gamma^{ab}C\right)  _{\alpha\beta}%
Z_{ab}\right]  ,
\end{align}
where we have used that $Z_{ab}=J_{ab,4}=\lambda_{4}\tilde{J}_{ab}$. \ This
procedure can be extended to any (anti)commutator of a $S$-expanded (super)algebra.

\bigskip

\section{Generalized Maxwell algebra in $D=4$ as an $S$-expansion}

\bigskip

\qquad In this appendix we will show how to obtain the generalized Maxwell
algebra $g\mathcal{M}$ from $\mathfrak{so}\left(  3,2\right)  $ using the
$S$-expansion procedure.

As in the previous cases, it is necessary to consider a subspaces
decomposition of the original algebra $\mathfrak{so}\left(  3,2\right)  $,%
\begin{equation}
\mathfrak{g}=\mathfrak{so}\left(  3,2\right)  =\mathfrak{so}\left(
3,1\right)  \oplus\frac{\mathfrak{so}\left(  3,2\right)  }{\mathfrak{so}%
\left(  3,1\right)  }=V_{0}\oplus V_{1},
\end{equation}
where $V_{0}$ \ is generated by the Lorentz generator $\tilde{J}_{ab}$ and
$V_{1}$ ist generated by the $AdS$ boost generator $\tilde{P}_{a}$.\ \ The
$\tilde{J}_{ab},$ $\tilde{P}_{a}$ generators satisfy the commutations
relations $\left(  \ref{ads01}\right)  -\left(  \ref{ads03}\right)  $, thus
the subspace structure may be written as%
\begin{align}
\left[  V_{0},V_{0}\right]   &  \subset V_{0},\label{subecs01}\\
\left[  V_{0},V_{1}\right]   &  \subset V_{1},\\
\left[  V_{1},V_{1}\right]   &  \subset V_{0}. \label{subecs03}%
\end{align}
Let $S_{E}^{\left(  2\right)  }=\left\{  \lambda_{0},\lambda_{1},\lambda
_{2},\lambda_{3}\right\}  $ be a finite abelian semigroup whose elements\ are
dimensionless and obey the multiplication law%
\begin{equation}
\lambda_{\alpha}\lambda_{\beta}=\left\{
\begin{array}
[c]{c}%
\lambda_{\alpha+\beta}\text{, \ \ \ \ when }\alpha+\beta\leq3,\\
\lambda_{3}\text{, \ \ \ \ \ \ \ when }\alpha+\beta>3.
\end{array}
\right.  \label{lm06}%
\end{equation}
Here $\lambda_{3}$ plays the role of the zero element of the semigroup
$S_{E}^{\left(  2\right)  }$. \ Let us consider a subset decomposition
$S_{E}^{\left(  2\right)  }=S_{0}\cup S_{1}$, with
\begin{align}
S_{0}  &  =\left\{  \lambda_{0},\lambda_{1},\lambda_{2},\lambda_{3}\right\}
,\\
S_{1}  &  =\left\{  \lambda_{1},\lambda_{2},\lambda_{3}\right\}  ,
\end{align}
This subset decomposition is said to be "resonant" because it satisfies
[compare with eqs.$\left(  \ref{subecs01}\right)  -\left(  \ref{subecs03}%
\right)  $.]%
\begin{align}
S_{0}\cdot S_{0}  &  \subset S_{0},\\
S_{0}\cdot S_{1}  &  \subset S_{1},\\
S_{1}\cdot S_{1}  &  \subset S_{0.}%
\end{align}

Imposing the $0_{S}$-reduction condition,%
\begin{equation}
\lambda_{3}T_{A}=0,
\end{equation}
we find a new Lie algebra generated by $\left\{  J_{ab},P_{a},Z_{ab},\tilde
{Z}_{ab},\tilde{Z}_{a}\right\}  $ where we have defined%
\begin{align}
J_{ab}  &  =J_{ab,0}=\lambda_{0}\tilde{J}_{ab},\\
P_{a}  &  =P_{a,1}=\lambda_{1}\tilde{P}_{a},\\
Z_{ab}  &  =J_{ab,2}=\lambda_{2}\tilde{J}_{ab},\\
\tilde{Z}_{ab}  &  =J_{ab,1}=\lambda_{1}\tilde{J}_{ab},\\
\tilde{Z}_{a}  &  =P_{a,2}=\lambda_{2}\tilde{P}_{a}.
\end{align}
These new generators satisfy the commutation relations%
\begin{align}
\left[  J_{ab},J_{cd}\right]   &  =\eta_{bc}J_{ad}-\eta_{ac}J_{bd}-\eta
_{bd}J_{ac}+\eta_{ad}J_{bc},\label{gm01}\\
\left[  J_{ab},P_{c}\right]   &  =\eta_{bc}P_{a}-\eta_{ac}P_{b},\label{gm02}\\
\left[  P_{a},P_{b}\right]   &  =Z_{ab},\label{gm03}\\
\left[  J_{ab},Z_{cd}\right]   &  =\eta_{bc}Z_{ad}-\eta_{ac}Z_{bd}-\eta
_{bd}Z_{ac}+\eta_{ad}Z_{bc},\label{gm04}\\
\left[  J_{ab},\tilde{Z}_{cd}\right]   &  =\eta_{bc}\tilde{Z}_{ad}-\eta
_{ac}\tilde{Z}_{bd}-\eta_{bd}\tilde{Z}_{ac}+\eta_{ad}\tilde{Z}_{bc}%
,\label{gm05}\\
\left[  \tilde{Z}_{ab},\tilde{Z}_{cd}\right]   &  =\eta_{bc}Z_{ad}-\eta
_{ac}Z_{bd}-\eta_{bd}Z_{ac}+\eta_{ad}Z_{bc},\label{gm06}\\
\left[  J_{ab},\tilde{Z}_{c}\right]   &  =\eta_{bc}\tilde{Z}_{a}-\eta
_{ac}\tilde{Z}_{b},\label{gm07}\\
\left[  \tilde{Z}_{ab},P_{c}\right]   &  =\eta_{bc}\tilde{Z}_{a}-\eta
_{ac}\tilde{Z}_{b},\label{gm08}\\
\text{others}  &  =0,
\end{align}
where we have used the multiplication law of the semigroup $\left(
\ref{lm06}\right)  $ and the commutation relations of the original algebra.
\ The new algebra obtained after a $0_{S}$-reduced resonant $S$-expansion of
$\mathfrak{so}\left(  3,2\right)  $ corresponds to a generalized Maxwell
algebra $g\mathcal{M}$ in $D=4$ \cite{19}. \ This new algebra contains the
Maxwell algebra $\mathcal{M}$ as a subalgebra. \ It is interesting to observe
that the $g\mathcal{M}$ algebra is very similar to the Maxwell algebra type
$\mathcal{M}_{6}$ introduced in Refs. \cite{8, 9}. \ In fact, one could
identify $Z_{ab}$, $\tilde{Z}_{ab}$ and $\tilde{Z}_{a}$ with $Z_{ab}^{\left(
1\right)  }$, $Z_{ab}^{\left(  2\right)  }$ and $Z_{a}$ of $\mathcal{M}_{6}$
respectively. \ However, the commutation relations $\left(  \ref{gm03}\right)
$, $\left(  \ref{gm06}\right)  \,\ $and $\left(  \ref{gm08}\right)  $ are
subtly different of those of Maxwell algebra type $\mathcal{M}_{6}$.

\end{document}